\documentclass[a4paper,10pt,preprint,5p]{elsarticle}

\usepackage{microtype}
\usepackage{hyperref}
\usepackage{tcolorbox}
\usepackage{enumitem}
\usepackage{graphicx}
\usepackage{xcolor}
\usepackage{amssymb}
\usepackage{chngpage}

\usepackage{listings}
\usepackage{xcolor}

\definecolor{codegreen}{rgb}{0,0.6,0}
\definecolor{codegray}{rgb}{0.5,0.5,0.5}
\definecolor{codepurple}{rgb}{0.58,0,0.82}
\definecolor{backcolour}{rgb}{0.95,0.95,0.92}

\lstdefinestyle{mystyle}{
    backgroundcolor=\color{backcolour},   
    commentstyle=\color{codegreen},
    keywordstyle=\color{magenta},
    numberstyle=\tiny\color{codegray},
    stringstyle=\color{codepurple},
    basicstyle=\ttfamily\footnotesize,
    breakatwhitespace=false,         
    breaklines=true,                 
    captionpos=b,                    
    keepspaces=true,                 
    numbers=left,                    
    numbersep=5pt,                  
    showspaces=false,                
    showstringspaces=false,
    showtabs=false,                  
    tabsize=2
}

\usepackage{tikz}
\usepackage{multirow} 
\usepackage{comment}
\usepackage{tabularx,booktabs,ragged2e,float,adjustbox,makecell}
\usepackage{tcolorbox}
\newcolumntype{L}{>{\RaggedRight\arraybackslash}X}
\newcolumntype{Y}{>{\Centering\arraybackslash}X}

\usepackage{subfig}
\usepackage{array}
\newcolumntype{R}[1]{>{\PreserveBackslash\raggedleft}p{#1}}

\journal{Journal of Software and System}

\usepackage{color,soul}

\usepackage{tikz}
\usepackage[english]{babel}
\usepackage{natbib}
\begin{document}

\begin{frontmatter}

\title{GitHub Copilot AI pair programmer: Asset or Liability?}

%\corref{c1}
\author{Arghavan Moradi Dakhel\corref{c1}} \author{Vahid Majdinasab\corref{c1}} \author{Amin Nikanjam, Foutse Khomh, Michel C. Desmarais}
\address{Polytechnique Montreal, Montreal, Canada}
\ead{\{arghavan.moradi-dakhel, vahid.majdinasab, amin.nikanjam, foutse.khomh, michel.desmarais\}@polymtl.ca}

\author{Zhen Ming (Jack) Jiang}
\address{%
  York University,
  Toronto,
  Canada}
\ead{zmjiang@cse.yorku.ca}  

%\author[a1]{Author1\corref{c1}}
%\address[a1]{Address: street, zip code \& city}
%\ead{author1@address.com}
\cortext[c1]{Corresponding authors. Both authors contributed equally to this research.}

% %% or include affiliations in footnotes:

% \author[mymainaddress,mysecondaryaddress]{}

% \author[mysecondaryaddress]{Global Customer Service\corref{mycorrespondingauthor}}
% \cortext[mycorrespondingauthor]{Corresponding author}

% \address[mymainaddress]{1600 John F Kennedy Boulevard, Philadelphia}
% \address[mysecondaryaddress]{360 Park Avenue South, New York}

\begin{abstract}
%\Foutse{the content of the letter doesn't seem to be synchronized with the text in the paper!}
Automatic program synthesis is a long-lasting dream in software engineering. Recently, a promising Deep Learning (DL) based solution, called Copilot, has been proposed by OpenAI and Microsoft as an industrial product. Although some studies evaluate the correctness of Copilot solutions and report its issues, more empirical evaluations are necessary to understand how developers can benefit from it effectively. In this paper, we study the capabilities of Copilot in two different programming tasks: (i) generating (and reproducing) correct and efficient solutions for fundamental algorithmic problems, and (ii) comparing Copilot’s proposed solutions with those of human programmers on a set of programming tasks. For the former, we assess the performance and functionality of Copilot in solving selected fundamental problems in computer science, like sorting and implementing data structures. In the latter, a dataset of programming problems with human-provided solutions is used. The results show that Copilot is capable of providing solutions for almost all fundamental algorithmic problems, however, some solutions are buggy and non-reproducible. Moreover, Copilot has some difficulties in combining multiple methods to generate a solution. Comparing Copilot to humans, our results show that the correct ratio of humans' solutions is greater than Copilot's suggestions, while the buggy solutions generated by Copilot require less effort to be repaired. %While Copilot shows limitations as an assistant for developers especially in advanced programming tasks, as highlighted in this study and previous ones, it can generate preliminary solutions for basic programming tasks. 
%\textcolor{blue}{R3-6: While Copilot  suggests solutions that are advanced in programming and comparable to humans in quality, it can become a liability if it is used by those who may not be familiar with the problem context and correct coding methods. Thus, alongside our suggestions on how to use Copilot more efficiently, an expert developer is still required to detect and filter its buggy or non-optimal solutions in software projects.}
Based on our findings, if Copilot is used by expert developers in software projects, it can become an asset since its suggestions could be comparable to humans' contributions in terms of quality. However, Copilot can become a liability if it is used by novice developers who may fail to filter its buggy or non-optimal solutions due to a lack of expertise.
\end{abstract}

\begin{keyword}
Code Completion, Language Model, GitHub Copilot, Testing.
\end{keyword}

\end{frontmatter}

%\linenumbers
%\begin{sloppypar}
\section{Introduction}
Recent breakthroughs in Deep Learning (DL), in particular the Transformer architecture, have revived the Software Engineering (SE) decades-long dream of automating code generation that can speed up programming activities. Program generation aims to deliver a program that meets a user's intentions in the form of input-output examples, natural language descriptions, or partial programs \cite{alur2013syntax, manna1980deductive, Gulwani2010}. 

Program synthesis is useful for different purposes such as teaching, programmer assistance, or the discovery of new algorithmic solutions for a problem~\cite{Gulwani2010}. One finds different approaches to automatic code generation in the literature, from natural language programming ~\cite{mihalcea2006nlp} and formal models \cite{drechsler2012generating,harris2016glast} to Evolutionary Algorithms \cite{sobania2021recent} and machine-learned translation \cite{rahit2019machine}.

Novel Large Language Models (LLMs) with the transformer architecture recently achieved good performance in automatic program synthesis \cite{brown2020language,chen2021evaluating,clement2020pymt5,feng2020codebert}. 
%\Foutse{talk a bit about diffrent code generation tools that have been proposed over the years before getting to codex!}
%Program synthesis is useful in different applications such as programmer's assistant, discovering new algorithmic solutions for a problem and automatic programming teaching~\cite{Gulwani2010}. 
One such model is Codex~\cite{chen2021evaluating}; a GPT-3~\cite{brown2020language} based language model with up to 12 billion parameters which has been pretrained on 159 GB of code samples from 54 million GitHub repositories.  Codex shows a good performance in solving a set of hand-written programming problems (i.e., not in the training dataset) using Python, named HumanEval dataset~\cite{chen2021evaluating}. This dataset includes simple programming problems with test cases to assess the functional correctness of codes. A production version of Codex is available as an extension on the Visual Studio Code development environment, named GitHub Copilot\footnote{\url{https://copilot.github.com/}}. Copilot, as an ``AI pair programmer'', can generate code in different programming languages when provided with some context (called prompt), such as comments, methods names, or surrounding code.

Several studies focus on the correctness of codes suggested by Copilot on the different types of problems such as linear algebra problems for an MIT course~\cite{drori2021solving} or university level probability and statistical problems~\cite{tang2021solving}. The author in~\cite{finnie2022robots} used Davinci (an API on a beta version of Codex) on different programming questions of a programming course and compared students’ grades with the grade of the tool in solving the programming questions correctly. There are few studies that assess other aspects of Copilot besides the correctness of its suggestions.~\citet{NguyenMSR22} compared the complexity of Copilot's solutions in different programming languages for several LeetCode questions, besides their correctness. Authors in~\cite{vaithilingam2022expectation} conducted a user study to understand how Copilot can help programmers complete a task. They studied how much time participants needed to complete a task using Copilot.

While these studies highlight some qualifications of Copilot, they neither examined the quality of the codes produced by Copilot compared to humans nor did they investigate the buggy solutions suggested by Copilot and the diversity of its suggestions.
Therefore, despite all these previous studies, we still do not know if--how Copilot, as an industrial component, can be leveraged by developers efficiently. We need to go beyond evaluating the correctness of Copilot's suggestions and examine how despite its limitations, it can be used as an effective pair programming tool.

The focus of our study is not on the type or difficulty level of programming tasks that Copilot can handle, but is on the quality of the codes that it will add to software projects if it is used as an AI pair programmer. We aim to investigate if the quality of codes generated by Copilot is competitive with humans and if it can be used instead of a developer in pair programming tasks of software projects without impacting code quality. We highlight Copilot’s limitations and competence with two different strategies and compared its suggestions with humans in different aspects. We also formulate suggestions on how developers can benefit from using Copilot in real software projects. 

%In this paper, we conduct an empirical evaluation of Copilot's strengths and weaknesses with two different strategies and formulate guidelines for its effective adoption as well as recommendations for its improvement. 

First, we assess Copilot's capabilities in solving fundamental algorithmic problems (i.e., searching and sorting) in programming. We study the correctness and reproducibility of Copilot’s solutions to these problems. Secondly, we compare Copilot's solutions with human solutions in solving programming tasks, to assess the extent to which it can mimic the work of a human pair programmer. We use a dataset of different programming tasks containing up to $4000$ in human-provided solutions (correct and buggy). % that gives us the opportunity to compare Copilot's solutions with them. 

To conduct our study, we have chosen datasets for which Copilot is able to generate answers corresponding to their programming tasks. While such tasks may not be representative of the typical programming tasks that a professional developer performs, they allow us to assess Copilot’s capabilities and limitations and to list our suggestions to developers on how to benefit from this tool in real software projects. We make the assumption that the findings on these datasets can generalize to more professional programming tasks and this is because ``programs, after all, are concrete formulations of abstract algorithms based on particular representations and structures of data''~\cite{wirth1985algorithms}.

The results of our study show that Copilot is capable of providing efficient solutions for the majority of fundamental problems, however, some solutions are buggy or non-reproducible. We also observed that Copilot has some difficulties in combining multiple methods to generate a solution. Compared to human programmers, Copilot's solutions to programming tasks have a lower correct ratio and diversity. While the buggy codes generated by Copilot can be repaired easily, the results highlight the limitation of Copilot in understanding some details in the context of the problems, which are easily understandable by humans.

Our finding shows Copilot can compete with humans in coding and even though it can become an asset in software projects if used by experts. However, it can also become a liability if it is used by novices, who may not be familiar with the problem context and correct coding methods. Copilot suggests solutions that might be buggy and difficult to understand, which may be accepted as correct solutions by novices. Adding such buggy and complex codes into software projects can highly impact their quality.

%The results of our study show that Copilot is capable of providing efficient solutions for the majority of fundamental problems, however some solutions are buggy or non-reproducible. We also observed that Copilot has some difficulties in combining multiple methods to generate a solution. Compared to human programmers, Copilot's solutions to programming tasks have lower correct ratio and diversity. While the buggy codes generated by Copilot can be repaired easily, the results highlight the limitation of Copilot in understanding some details in the context of the problems, which are easily understandable by humans. 

To summarize, this paper makes the following contributions:

\begin{itemize}
    \item We present an empirical study on the performance and functionality of Copilot's suggestions for fundamental algorithmic problems.  
    \item We empirically compare Copilot's solutions with human solutions on a dataset of Python programming problems.
    %\item We indicate and discuss reproducibility of the  results when experimenting with Copilot. 
    \item We make the dataset used and the detailed results obtained in this study publicly available online \cite{GithubRepo} for other researchers and--or practitioners to replicate our results or build on our work.
    % \item We shed light on the 
\end{itemize}

\textbf{The rest of this paper is organized as follows}. We briefly review the related works in Section~\ref{sec:related}. Section~\ref{sec:methodology} presents the design of our study to evaluate Copilot as an assistant to developers. We report our experiments to assess Copilot's suggestions for fundamental algorithmic problems and compare generated suggestions with what programmers do on specific programming tasks in Section~\ref{sec:results}. We discuss our results and potential limitations in Section~\ref{sec:discussion}. Threats to validity are reviewed in Section~\ref{sec:threats}. Finally, we conclude the paper in Section~\ref{sec:conclusion}.
%\Foutse{why do you have these in separate sections? it sounds like you are having two papers patched together!!! please try to have a common methodology section for your paper! same for analysis and results!}

%Research in Software engineering on the automated generation of source code\\
%GitHub's Copilot\\
%\section{Background}
%Perhaps on language models,\\
%and on the GitHubs's Copilot and how it works\\
%and any other related background
%%%%%%%%%%%%%%%%%%%%%%%%%%%%%%%%%%%%%%%%

\section{Related Works}\label{sec:related}
%In this section, we review the literature... 
%\subsection{Program Synthesis}
%Program synthesis is the process of generating a program that meets users' intention. The users' intention can be specified in a form of input-output examples, natural language description or partial program~\cite{alur2013syntax, manna1980deductive}. Program synthesis is useful in different applications such as automatic teaching, programmer's assistant or discovering a new algorithmic solutions for a problem~\cite{gulwani2010dimensions}. In recent years, with the spread of deep learning, program synthesis tools have also been influenced by this technique. \citet{yin2017syntactic} proposed a RNN model that generates Abstract Syntax Tree (AST) for a text and then convert the AST into the code in Python. 
%\subsection{Language Models}
%\subsection{Copilot}
%\Foutse{use the gaps in these analysis to motivate your work?}
A few studies empirically investigate the different capabilities of Copilot. \citet{sobania2021choose} compared Copilot with a Genetic Programming (GP) based approach that achieved good performance in program synthesis. Their findings show that GP based approaches need more time to generate a solution. Moreover, training GP based models are expensive due to the high cost of data labeling. % hand-labeling training dataset. 
Also, these approaches are not suitable to support developers in practice as GP usually generates codes that are bloated and difficult to understand by humans \cite{sobania2021choose}.

\citet{vaithilingam2022expectation} conducted a human study involving $24$~participant to understand how Copilot can help programmers to complete a task. They focused on $3$ Python programming tasks: ``1. edit CSV, 2. web scraping'' and ``3. graph plotting''. Their finding shows that while Copilot did not necessarily improve the task completion time and success rate, programmers prefer to use Copilot for their daily tasks because it suggests good starting points to address the task. The tasks in this study involve less problem solving effort compared to the typical programming tasks in our study.  They are mostly related to using programming language libraries. Also, they did not compare Copilot's suggestions with their participants' suggestions when working without the help of Copilot.

\citet{drori2021solving} studied Copilot's capability in solving linear algebra problems for the MIT linear algebra course. In the same line of work, Tang et al. examined Copilot's capability in solving university level probability and statistical problems~\cite{tang2021solving}. These two studies only focused on the correctness ratio of Copilot's solutions and did not examine its performance on programming tasks.

 ~\citet{finnie2022robots} used Davinci (an API on a beta version of Codex) on two datasets. The first dataset includes 23 programming questions for a programming course, students' solutions for these questions, and their grades. This dataset is not publicly available. The second dataset is a set of different descriptions of a single well-known problem, rainfall, without humans’ solutions. For the programming questions, the paper focused on the grading of the solutions suggested by Codex: generating the correct solution for the problems after different runs (10 runs) and then comparing the grading with students. For the second dataset, besides the code correctness, they checked the variety of solutions by calculating the number of source lines of code (sloc). Their results showed that Codex outperformed most students as evidenced by the grades received for their proposed solutions. Also, they observed that using the same input as a prompt on Codex can lead to different solutions, while Codex can generate correct solutions for different descriptions of the same problem.

\citet{NguyenMSR22} evaluated Copilot on $33$ LeetCode questions in $4$ different programming languages. They used the LeetCode platform to test the correctness of Copilot's solutions. The questions in their study included different levels of difficulty. Although they evaluated the correctness of Copilot's solutions and compared their understandability, % of Copilot's suggestions in these $4$ programming languages. However, 
they did not assess whether Copilot successfully found the optimal solution for each task.

Another group of studies focuses on vulnerability issues to evaluate Copilot solutions. As mentioned before, Copilot is trained on a large volume of publicly available code repositories on GitHub which may contain bug or vulnerability problems.
%\Foutse{please rephrase the following section to make sure to identify gaps in the studies and connect with your work!}
\citet{pearceasleep} conducted different scenarios on high-risk cybersecurity problems and investigated if Copilot learns from buggy code to generate insecure code. Another study investigates how Copilot can reproduce vulnerabilities in human programs~\cite{asare2022github}. To do so, they first used a dataset of vulnerabilities generated by humans, then rebuilt the whole code before the bug %\Ahura{this makes no sense} 
and asked Copilot to complete the code. The completed section was manually inspected by three coders to determine if Copilot reproduced the bug or fixed it.

\citet{moroz2022potential} examined the challenges and the potential of Copilot to improve the productivity of developers. They highlighted the copyright problems and the safety issues of its solutions. They discussed the non-deterministic nature of such models and the harmful content that could be generated by them.

Authors in~\cite{ziegler2022productivity} surveyed 2631 developers about the impact of Copilot on their productivity and highlighted different metrics of users' interaction with Copilot that help predict their productivity. They relied on the SPACE~\cite{forsgren2021space} framework to generate 11 Likert-style questions in their survey. Also, they analyzed the usage data of Copilot of the participants who responded to this survey. They extracted different metrics from this data such as the acceptance rate of solutions, persistence rate, unchanged and mostly unchanged accepted solutions, etc. They found that the acceptance rate of solutions by developers is the most relevant metric that shows the impact of Copilot on the productivity of developers.
%\Foutse{what did they found?}
 
To the best of our knowledge, none of these studies compared Copilot with humans for solving programming tasks. The majority of these studies focused on assessing the correctness of Copilot's solutions and highlighted its issues; e.g., the presence of vulnerabilities in generated solutions.  %\Ahura{either as a vulnerability or vulnerabilities}. 
In this study, we focus on fundamental algorithmic problems and compare Copilot with humans in solving real world programming tasks.  
%We highlight its strength over humans \Foutse{how is this highlighting its strength over humans? please clarify!} code by discussing the efforts required to repair Copilot's buggy solutions with a repairing tool and the complexity of its suggestions.

%\Amin{we also need to mention:}Copilot's similar performance to Genetic Programming in standard program synthesis benchmarks \cite{sobania2021choose}, too complex codes and relying on undefined helper methods for particular tasks \cite{NguyenMSR22}.
%Another study investigated if different natural language descriptions that are semantically equivalent can caused different code recommendations~\cite{}\Arghavan{it is not published yet ``On the Robustness of Code Generation Techniques:An Empirical Study on GitHub Copilot''}\Amin{so, we can't cite it!}. They used a DL-based paraphrasing technique to generate semantically equivalent descriptions. Their results shows the recommended code changed in 70\% of the cases.
%\Amin{we need to highlight the difference or similarity of our work to others}

\section{Study Design} \label{sec:methodology}
%\Foutse{we need a common methodology section where we motivate the two rqs, explain how their combined results help address our main goal, before going on and explaining how we conduct analysis to answer the Rqs!}
%\Amin{@Foutse, I revised, please have a look}
In this section, we present our methods to assess Copilot as an AI pair programmer and detail the experimental design to achieve our goals. 

%\textcolor{blue}{R-1-1: To assess Copilot as an ``AI pair programmer'' we perform a two-steps empirical study based on different quantitative metrics of code quality and correctness on solutions suggested by Copilot for various programming tasks. However, the quantitative metrics that we are reporting in this study, such as optimality, complexity, or repairing cost, are not calculated in the manner of calculating developers' contributions. Still, these are important metrics to evaluate the quality and effectiveness of their contributions.} 

%\textcolor{blue}{Due to Copilot's limitation in understanding complex programming tasks, we select the tasks at a level of difficulty that Copilot can be responsive to, thus we can apply our quantitative analysis further. Our study is a step in assessing the benefits and boundaries of using Copilot as an AI pair programmer in software projects.}
%we need to see how it can effectively assist developers during the development process. 

To solve a programming task, a developer usually builds the code on top of fundamental data structures (e.g., queues, trees, graphs) and algorithms (e.g., search, sorting) in computer science. %, like and search/sort algorithms. 
%\Foutse{what do you mean by having hand? please we better use a clear wording !!!} So, having hands on such structure/algorithms is necessary for a developer. 
Moreover, the developer needs to come up with ideas %\Foutse{would copilot help with creativity?}
to achieve the goal(s) of the programming task efficiently. 
%As the development is an iterative process and a developer may examine her code several times for refinement/debugging, the bug-proneness and creativity of automatically generated solutions should be assessed. Therefore, an ideal automatic code generator should be able to recommend proper structure/algorithms and innovative clues to the developer, so that the developer can build on them. The recommended code is expected to be competitive with what humans may generate during the development.

%As none of the previous studies examined the effectiveness of Copilot as a programming assistant, 
We evaluate Copilot on 1) the adequacy of recommended code for fundamental algorithmic problems, and 2) the quality of recommendations compared to human-provided solutions. Specifically, we address the following research questions (RQs): %Our Research Questions (RQs) in this paper are as follows: %investigate reproducibility of Copilot’s outputs,  Therefore, our Research Questions (RQs) in this paper are as follows:

\begin{itemize}
    \item [\textbf{RQ1:}] Can Copilot suggest correct and efficient solutions for fundamental algorithmic problems?
    %\begin{itemize}
    %\item{Are Copilot's suggestions for these problems reproducible?}
    %\end{itemize}
    \item [\textbf{RQ2:}] Are Copilot's solutions competitive with human solutions for solving programming problems?
        %\begin{itemize}
        %\item{How difficult is it to repair Copilot's buggy suggestion compared to human solutions?}
    %\end{itemize}
    %\Foutse{add a brief motivation for this question and a brief summary of the key findings!}
\end{itemize}

%In the first strategy,  \Foutse{why is this important to know? what is the motivation for selecting this specific task?} \Ahura{I added some more explanation. please take a look}.

\begin{figure*}
\centering
 \includegraphics[width=0.95\linewidth]{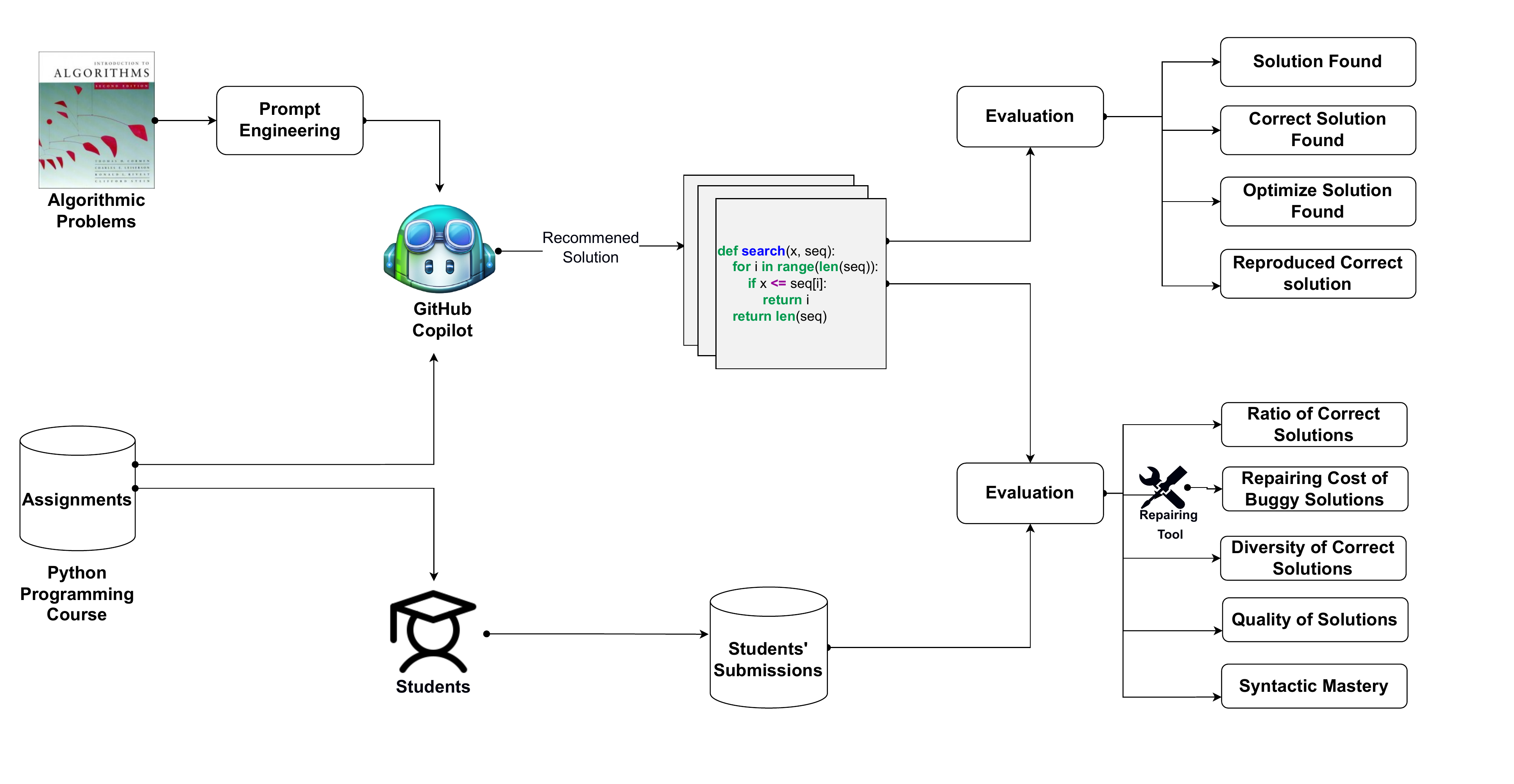}
  \caption{\textbf{The Workflow of proposed methods.} The study includes two different methods to test Copilot in recommending codes to solve programming problems. The first pipeline focuses on algorithmic problems collected from a well-known algorithm design book~\cite{cormen2022introduction}. The second pipeline focuses on the assignments of a Python programming course~\cite{hu2019re}. It compares Copilot with students in solving programming problems in different aspects.}
  \label{fig:overview}
\end{figure*}

In the rest of this section, we describe the research methods we followed to answer each of our RQs as illustrated in Figure~\ref{fig:overview}.

\subsection{\textbf{RQ1: Copilot on Algorithm Design}}\label{sec:RQ1_all}

    Our goal in RQ1 is to observe if Copilot can generate solutions for fundamental algorithmic problems given clear descriptions of the problem and do further analysis.  Solving these fundamental algorithmic problems is important for developers contributing to a software project. Although these problems are not necessarily typical of professional projects, understanding how to solve them with an optimal algorithm is essential because professional programming tasks are a combination of fundamental algorithmic and data structure problems~\cite{wirth1985algorithms}. Hence, the ability to address basic programming problems correctly and efficiently is a prerequisite for addressing more professional programming tasks. Further, failing to perform them is likely to translate into failure on more complex software engineering tasks. In this section, we plan to examine if Copilot is capable of solving these fundamental problems with optimal solutions.  %Although these problems are not very practical, but these algorithmic problems include in the technical assessments of programmers to assess their coding ability. For example, Leetcode\footnote{\url{https://leetcode.com/}} is a website that represents different categories of such problems for the coding assessment. Developers may not use these algorithmic problems outside of the assessment (i.e. in their daily tasks), but understanding how to solve them with an optimal algorithm is essential. In addition, it gives us the possibility to test Copilot’s ability in code generation from problem descriptions written by humans and assess its suggested solutions with different quantitative metrics, since Copilot is responsive to the majority of these selected tasks.
    %(the same way developers are assessed in their coding ability). 
    
    %The algorithms we have chosen are fundamental in software development. Therefore, it is imperative to study Copilot's ability in generating code for them since it claims to take the position of a pair programmer. %\Foutse{can we elaborate on the implication of the result? for example explaining that if copilot can achieve a good performance on fundamental algorithm it can be leveraged in xxx way, because fundamental algorithms are the cornerstone of yYY  possibly providing some references!}
    \subsubsection{\textbf{Data set: Fundamental Algorithmic Problems}}
    
        We selected the problems and their descriptions from \cite{cormen2022introduction}. We choose this resource because it is %a highly cited and prestigious textbook that is 
        widely used for teaching algorithmic design fundamentals to computer science students \cite{cormen2022citations}. %\Foutse{XXX}
        In this book, the authors explain the principal algorithms that computer engineers must be knowledgeable about by breaking them into categories. Since our problems were selected from this book, we followed its categorization, such that our tests on Copilot were conducted on 4 categories:
        
        \begin{itemize}
            \item \textbf{Sorting Algorithms:} Sorting algorithms are among the first algorithmic concepts that are taught to computer science students. These algorithms introduce the concept of time complexity and how inefficient code can make differences in more complex programs. Sorting algorithms are used in databases to segment large amounts of data that cannot be loaded entirely into memory or in numerical operations to determine which numbers should undergo operations first for the results to be produced as quickly as possible.
            From this section, we selected some well-known sorting algorithms which students are asked to learn and then implement. These algorithms are methods for sorting an array of numbers (integers or floats) in a descending or ascending manner. In these problems, time complexity, a measure of an algorithm's run-time as a function of input length, is an important factor to be considered.
            
            From the algorithms described in the book, we selected \textit{bubble sort}, \textit{bucket sort}, \textit{heap sort}, \textit{insertion sort}, \textit{merge sort}, \textit{quick sort}, \textit{radix sort}, and \textit{selection sort}. We selected these algorithms based on their implementation complexity, from easy to hard, based on \cite{cormen2022introduction}'s descriptions. %\Foutse{assuming copilot succeed on these algorithms, how can developers use it in a real software development context? we need to make the connection i think!}
            %Copilot's ability to understand and reproduce code for these algorithms, will allow the programmer to use the generated code in their code bases and instead focus on more complex parts of the program.
            %\Amin{shouldn't be in results?}
            
            \item \textbf{Data Structures.} From this section, we selected the Binary Search Tree (BST). BST is a basic data structure that is taught in algorithm design. Each node of the tree contains a value (called \textit{key}) that is greater than all the values in the left sub-tree and smaller than all the values in its right sub-tree. The implementation of BST involves multiple steps, namely:
            
            \begin{itemize}
                \item Finding the minimum and maximum values in the tree before inserting any new value.
                \item In-order tree walks to extract all the values in the tree in a sorted manner.
                \item Finding the successor node. Given a node \textit{x}, the successor of \textit{x} is the node that has the smallest value which is greater than \textit{x}.
            \end{itemize}
            %\Foutse{similar how copilot success on this can be leverage in software development tasks?}
            
            \item \textbf{Graph Algorithms.} From this section, we selected the \textit{Elementary Graph Algorithms}. These algorithms are used to perform some elementary operations on a graph. %\Foutse{beyond knowledge testing, how does performing well on these tasks can be leverage in software development? we need to make the connections i think!} 
            Since graphs store information about how each node is connected to others, they can be used in implementing applications such as maps and social media user connections.
            We tested Copilot on the following graph algorithms problems:
            %Graph are comprised of a set of nodes and vertices. Each node can be connected to multiple nodes through vertices. 
            \begin{itemize}
                \item Generating code for a simple graph data structure.
                \item Breadth First Search (BFS) on a graph.
                \item Depth First Search (DFS) on a graph.
                \item Implementing Directed Acyclic Graphs (DAG). DAGs require a more complex implementation logic compared to simple graphs, since during initialization, based on the directions of the edges, we need to check if a cycle exists in the graph or not. %Our goal with this task is to see whether Copilot can understand and generate code for this task or not.                % \Michel{what is the task exactly?}                  
                \item Finding reachable \textit{vertices}. A pair of vertices are defined as reachable if both vertices can be reached from each other in a directed graph. %Here, the task is to check if two vertices can be reached from \textit{each other}.
                % \Michel{again, what is the task exactly?  finding cliques or determining if a graph is strongly connected?}
            \end{itemize}
            
            \item \textbf{Advanced Design and Analysis Techniques.}  We selected the greedy algorithms from this section. Unlike the algorithms described above, the greedy technique is a general approach for solving optimization problems based on breaking problems down into multiple subproblems and selecting the best solution at the given time. As these solutions need to be evaluated in the context of a problem, we selected the ``activity selection'', an introductory problem to greedy algorithms as described in \cite{cormen2022introduction}.
            %from \cite{cormen2022introduction}.
        \end{itemize}
        
    \subsubsection{\textbf{Prompt Engineering}}\label{sec:prmtEng}
        Alongside code completion, Copilot can generate code from natural language descriptions in the form of comments. However, as noted by \cite{li2022competition}, if the description becomes too long or detailed, Copilot's performance degrades. Since the book that we are using to collect the problems \cite{cormen2022introduction} is a comprehensive educational book, %the authors assumed that the reader has no experience in algorithm design. For this reason, 
        each problem is described in detail and by building upon concepts that were explained in the previous chapters. As a result, some problem descriptions span multiple paragraphs and sometimes, pages.

         However a summary description of our selected problems can be found in different resources, but the authors summarized the description of each problem in their own words to reduce the chance of memorization~\cite{carlini2022quantifying} issue on Copilot. Therefore, our prompt engineering was done in two steps:
        \begin{enumerate}
            \item \textbf{Describing the problem:} We needed to summarize each problem's description to feed them to Copilot while staying as faithful as possible to the books. To make sure that our descriptions were understandable and did contain enough information about the algorithm being described, we cross-checked each of them with those on popular coding websites such as W3SCHOOLS~\cite{w3schools} and GEEKSFORGEEKS~\cite{geeksforgeeks} as well. For cross-checking, the second author %\Amin{Who? the first two authors? and how? 2 or 3 raters did this!like: the first authors summarize, then the second author check them. if there is any conflict,they meet and discuss \Ahura{I understand. I'll re-write this part after the our review is done. Instead of writing and re-writing it over and over again, I want to do it in one go}}
           summarized \cite{cormen2022introduction}'s algorithm descriptions while keeping in mind Copilot's limits on the length of the prompt. If there were differences in the descriptions (i.e., the description was missing some key elements in explaining the problem), the descriptions were revised.%\Foutse{who is we? was this done by multiple people? did people always agreed? i think we need to be more precise on the generation of prompts! if more people participated, we need to compute kappa agreements} \Ahura{We will edit this after we calculate a kappa score} revised our description. %while keeping Copilot's limits in mind\Amin{not be very long?}. 
            %were revised. This approach was taken for sorting algorithms, binary search trees, and the ``activity selection'' problem.  
            
            %\item \textbf{Asking Copilot to generate a solution by mentioning Algorithm/Problem's name:} For elementary graph algorithms, instead of describing the problem as if the developer has no knowledge about the algorithm, we asked Copilot to generate codes by directly asking what we were looking for. %\Foutse{which development scenario does this correspond to!}.
            %For example, instead of describing what a graph is, we asked "\textit{create a class that implements a graph data structure}". We did this for two reasons:
            %\begin{itemize}
               % \item The graph data structure is well known even to novice programmers and we are trying to assess Copilot's ability in generating code in the same level of novice and intermediate programmers.
              %  \item Given Copilot's position as a pair programmer, it would be counter productive to explain fundamental algorithms during software design and production. Hence, we wanted to see if Copilot is able to recognize well-known data structures without having to describe what they are in-depth. %\Foutse{what would be the use case for this? can we explain?}
             \item  \textbf{Cross validation of problem descriptions:} Cross-validation of problem descriptions: The second author created the input descriptions as explained above. After this, these descriptions were cross-checked with the first author to make sure that they were correct, understandable, and contained enough information about the problem being described. The first two authors both have taken the course “Introduction to Algorithms” during their education and have more than 5 years of experience in coding and program design. To assess the agreement, we have calculated Cohen's Kappa score \cite{CohenKappa}. While the score was 0.95 implying an \textbf{almost perfect agreement}, for cases where there were conflicts about the descriptions, the two authors met and discussed the conflicts to resolve them. In the end, the descriptions were also cross-checked with the third author who has more than 10 years of experience in teaching algorithm design. Therefore, the final input descriptions were what all three authors agreed on. 
        \end{enumerate}
Excluding sorting algorithms, other problems require code to be generated using previous code as it is common practice in both functional and object oriented programming %\Amin{you may add a reference \Ahura{I don't think it needs one!}}. %\Foutse{can we tied this to a concrete software development scenario?}.
For these problems, we followed exactly the book's example by asking Copilot to generate code for the underlying subproblems %\Amin{subproblem? \Ahura{I think concept is better}} 
and then for the succeeding problems, we asked it to implement the solution using the code it had generated before. We have recorded all of our descriptions and Copilot's responses in our replication package \cite{GithubRepo}.  
           % \end{itemize}
        
\subsubsection{\textbf{Solving Fundamental Algorithmic Problems with Copilot}}\label{subsec:cop_algo_data}
        
 To generate solutions with Copilot, we feed the description of each algorithmic problem, call it prompt, to Copilot. At each attempt on Copilot with a selected prompt, it only returns up to the top $10$ solutions. Thus, we do not have access to the rest of the potential suggestions. To inquire about Copilot's consistency in generating correct solutions, we repeat the process $6$ times and each time collect its top $10$ suggestions.

To assess whether Copilot's suggestions are consistent over time, we performed 2 trials within a 30 days time window. Each trial consists of $3$ attempts for each prompt and each attempt contains up to $10$ suggestions provided by Copilot. The collection of $3$ first attempts is called ``First Trial'' and the collection of $3$ last attempts which were conducted $30$ days later is named ``Second Trial''.

Given that Copilot may try to consider the script's filename as a part of its query, to make sure that solutions were only generated from our descriptions, we gave the scripts unrelated names.

        % For both approaches, the descriptions were checked by both the first and second authors to make sure that they were understandable, contained enough information about the problem being described, and correct. In order to reach a consensus, the conflicts were discussed with the third author and Kappa scores were calculated for each problem. 

    \subsubsection{\textbf{Evaluation Criteria}} \label{Sec:algo_eval}
        %Since our problems are well known between computer engineers, and since \cite{cormen2022introduction} does not contain code samples for each problem, both authors checked the results of code generation, and calculated the Kappa agreement between them. 
        %From each category, we selected a problem and asked Copilot to generate code for solving it. 
        Below, we briefly explain the $4$ different metrics which we have used to evaluate Copilot and explain them in detail in the rest of this section.  The metrics are calculated per each fundamental algorithmic problem.
        
        \begin{enumerate}
            \item\textbf{Response Received $\in [0,3]\in \mathbb{N} $.}
            The number of attempts in each trial that Copilot was able to understand the problem and generate code content as its response.
            \item \textbf{Correct Ratio (\%).} The percentage of correct solutions suggested by Copilot in each trial.  %Copilot presents its suggestions in a list so that the developer can choose which of the solutions can be accepted as correct code. For this marker, we consider both syntactical and operational correctness of the suggested codes. Therefore, for evaluating correctness, we check whether the generated codes run without any errors and whether they contain bugs after being compiled. %\Amin{compile without error and produce the intended output}.
            %If out of the 10 suggestions that Copilot produces in an experiment, even one of them is correct, we consider it as a successful attempt.
            %\Foutse{why? do we assume that a paired developer will be able to recognise the good solution?}
            \item \textbf{Code Optimality $\in [Yes, No]  $}. Whether at least one of the correct solutions suggested by Copilot in each trial has optimal time complexity [Yes or No].
            
            \item \textbf{Code Reproducibility $\in [Yes, No]  $.}
            \begin{itemize}
                \item \textbf{Within a Trial:} Whether at least one of the correct solutions suggested by Copilot in one attempt was repeated in two other attempts, within a trial [Yes or No].
        \item
            \textbf{Across Trials:} Whether at least one of the correct solutions suggested by Copilot in the first trial was repeated in the second trial   [Yes or No].
            \end{itemize}
            
             %\Foutse{consistency or reproducibility?} \Ahura{tbh, both! problem is that copilot does some variable name changes and some structure changes. that's why we did AST analysis.}
    \item 
    \textbf{Code Similarity $\in [0,1]\in \mathbb{R} $.} 
                \begin{itemize}
                \item \textbf{Within Trial:} The similarity degree between all correct solutions within a trial.
        \item
            \textbf{Across Trials:} The similarity of correct solutions between two trials.
            \end{itemize}

        \end{enumerate}
    
       %\Foutse{this sentence is a repetition!} All the data we collected during this study can be accessed at our replication package \cite{GithubRepo}.
       %\Foutse{query execution? we need to put this in a section with a title!} \Ahura{I don't understand what you mean.}\Amin{you can revise the title of this subsection to "query execution and evaluation metrics"}
      \subsubsection*{\textbf{(1) Response Received}}
       
       Our observation shows that if Copilot is unable to provide solutions to the problem with the provided prompt, it will return irrelevant responses such as repeating user's prompts, code that only contains import statements or natural language responses. Thus, this metric helps us to evaluate if Copilot generates code for the summarized description of the program instead of the mentioned irrelevant responses.

        We used the description of each problem in the form of comments and collected up to the top 10 suggestions of Copilot in 6 different attempts and two separate trials, as it is described in Subsection~\ref{subsec:cop_algo_data}. To calculate this metric, if at least one of the suggested solutions in an attempt within a trial is code content, we consider it as a successful code generation attempt or ``Response Received''. Since we conduct 3 separate attempts in each trial, we report the value of this metric with a number $\in [0,3]\in \mathbb{N} $ per trial. 
        %Then, Copilot was asked to generate code for the input description. Given that Copilot tries to understand the general purpose of the code from the script's filename%\Amin{script of filename?}
        %, to make sure that solutions were generated from our descriptions, we gave the scripts unrelated names. For example, if we were testing on bubble sort, we named the corresponding script as \textit{"script\_1"}. 

        \subsubsection*{\textbf{(2) Correct Ratio}} \label{sec:algo_correct}
        
We report the correct ratio as a fraction of solutions suggested by Copilot per problem that are functional and address the objective of the problem. To calculate this metric, we first need to evaluate the correctness of Copilot's suggestions for a problem. A suggested code is correct if it passed a set of unit tests. 

However, in algorithmic problems, passing a set of unit tests to check the correctness of solutions is not enough. In this category, not only we need to verify a suggestion on passing a set of unit tests, but also we need to verify its chosen algorithm. 

For example, in the ``Sorting'' problems, all problems have the same functionality: sorting a list of numbers. But the importance is the choice of the algorithm to address the problem and to check if Copilot is able to understand the structure of the solution from the given description. If Copilot implements the ``Bubble sort'' instead of the ``Selection sort'' algorithm or uses the Python built-in functions ``sort'' or ``sorted'', the code is still functionally correct and is able to sort the inputs. But the code is not addressing the algorithm described in the problem. That is the same situation for implementing the data structure of a BST or a graph. 

We tackle this challenge of calculating the correct ratio by following three steps:
\begin{enumerate}
\item We check the functional correctness of Copilot's suggestions on a set of unit tests.
\item We check if the selected algorithm in the solution follows the description that we gave to Copilot for that problem. To conduct this step, same as in Subsection~\ref{sec:prmtEng}, the two first authors separately checked the solutions suggested by Copilot for the problems. They compared the algorithm of the solutions (that is employed by Copilot to solve the problem) to the reference algorithms (ground truth). We collect the ground truth for each problem from the reference book~\cite{cormen2022introduction} and from popular coding websites such as W3SCHOOLS~\cite{w3schools} and GEEKSFORGEEKS~\cite{geeksforgeeks}. We calculate Cohen's Kappa score to measure the agreement between the two authors.
\item The solutions per problem within a trial that passed the two above steps are labeled as correct. Then, we calculate the correct ratio based on the fraction of the correct solutions within a trial.
\end{enumerate}

\subsubsection*{\textbf{(3) Code Optimality}}

 We report this metric because the problems in our dataset can be implemented with different algorithms. This choice of the algorithm may impact their computation complexity for example using a nested loop, queue, or recursive functions to solve a problem. With this metric, we want to check if Copilot is able to suggest the optimal algorithm of a problem among its correct suggestions.

 We cannot write a code to automatically check if the computation size of another code is optimal due to Turing’s halting problem~\cite{bera2020fundamental}. Thus, same as in Subsection~\ref{sec:prmtEng} and Correct Ratio in this section, the two first authors check if there is a solution with an optimal algorithm between the correct solutions suggested by Copilot for a problem in a trial. They separately compared correct solutions with a reference optimal code for a problem (ground truth). If at least one of the correct solutions suggested by Copilot within a trial is optimal, they consider that Copilot is able to find an optimal solution for that problem [Yes] and otherwise [No]. We calculate Cohen’s Kappa score to report the agreement of two authors on code optimality. 

%\textcolor{blue}{We select this metric because the algorithm problems that we select can be implemented with different code i.e. nested loop, using queue, recursive method, etc. 
            % For testing code correctness and optimality, we took 2 steps:
            %\begin{itemize}
               % \item In order to %calculate a Kappa score and 
               % analyze generated codes' time complexity, the codes were inspected manually by the first two authors.
               % \textcolor{blue}{R-1-10: It should be noted that optimality analysis was only done on codes that were considered to be correct. As codes that have syntax errors cannot be run and codes that are simply incorrect cannot be compared to correct codes in terms of optimality.}
                %\Amin{both?!do you mean the first two authors?}.%\textit{\textbf{by the second author}} \Foutse{by who?}. %by both authors.
                
               % \item In order to make sure that the generated codes were correct and functioned as intended, we wrote unit tests for each script and ran them to make sure that the generated solution runs without any syntax errors and/or bugs. All scripts are accessible in our replication package.%\Amin{are they available in the replication package? if yes mention it}.
            %\end{itemize}

    %\Foutse{we need a sub section here too!}
   % \paragraph{\textbf{Testing Code Reproducibility:}}
\subsubsection*{\textbf{(4) Code Reproducibility and Similarity}}        

 While Copilot is closed-source and we have no information about its characteristics that may impact its behavior on our prompts, we want to study if this tool is able to reproduce a correct solution for a problem in different attempts and over time. We introduce ``Code Reproducibility'' as a metric for this measurement. For more clarification, we split our approach for measuring this metric into three subsets:
\begin{itemize}
    \item We consider two different types for reproducing a code: the one that checks if a correct solution is reproduced across different attempts within a trial and calls it ``Within a Trial'', and the one that checks if a correct solution of a problem is reproduced over a time window among two trials and call it ``Across Trials''. 

    \item  To identify the correct solutions that are reproduced and measure their similarity, we have used the Abstract Syntax Trees (AST) similarity method described in \cite{salazar2020comparing}. AST similarity is calculated by first building the AST of a code and then pruning the leaves that are related to variable or function names. Also, we ignore comments or any natural language text in each solution as they are not part of the code itself.

    AST similarity is bounded between 0 and 1 with 1 denoting structurally equivalent programs (regardless of their semantic similarity) and 0 denoting no equivalence between programs. It also returns 1 for “structurally equivalent recorded programs” where the programs are functionally identical but their instructions are executed in a different order, and “structurally equivalent renamed identical programs” where the programs are structurally the same with different variable names.

    Therefore, this similarity measure will not be affected by different statement orders or different variable names. However, this similarity will be different for semantically similar programs where the same concept is implemented in different ways. In Subsection~\ref{sec:criteria}, we explain in more detail why we need to apply this method to detect similar codes when we discuss Copilot's duplication solutions.

    \item  To apply this comparison to correct solutions ``Within a Trial'', we compare the pairs of correct solutions across 3 different attempts within that trial. If at least one of the correct solutions in one attempt is reproduced in two other attempts (similarity equals 1), or in other words if at least one of the correct solutions within a trial occurs in all its 3 attempts, we consider that Copilot is able to reproduce the correct solution for that problem ``Within a Trial'' [Yes], otherwise, we consider that [No]. To apply it  ``Across Trials'', we compare the pairs of correct solutions among two trials. If at least one of the correct solutions in the first trial is reproduced in the second trial (similarity equals 1), we consider that Copilot is able to reproduce the correct solution for that problem ``Across Trials'' [Yes], otherwise, we consider that [No].
\end{itemize}

        %For testing code reproducibility, first we need to ensure that our results are as independent from each other as possible. So, we terminated our internet connection and closed the editor after each experiment. After which, we re-connected to the internet, opened the editor, and asked Copilot to generate solutions by giving it the exact same description. Another step in testing reproducibility  For quantifying the reproducibility between trials, we have compared the Abstract Syntax Trees (AST) of each of Copilot's suggestions between each test. To do this, we have used the code %\Amin{approach?method?}
        
\begin{figure}
\centering
  \includegraphics[width=0.85\linewidth]{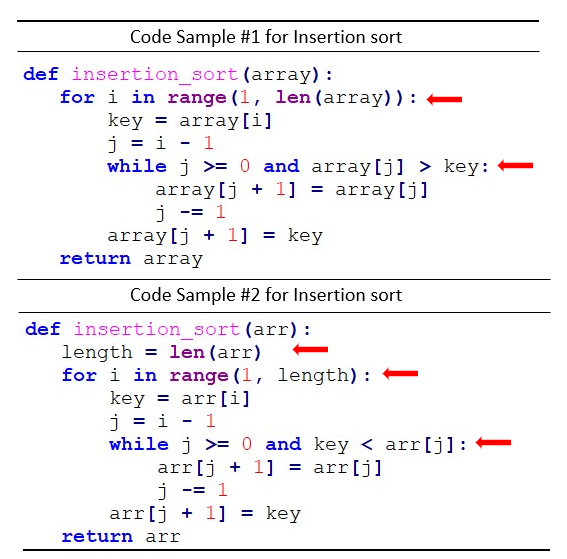}
  \caption{\textbf{ Two different solutions suggested by Copilot for Insertion sort. There are a few lines in these two code samples that are syntactically different but both are addressing the same functionality.} Code Sample \#1 calculates the length of the list within the range in for loop instructor. Code Sample \#2 assigns the length of the list to a variable and then uses it to control the loop. The comparison operator in the while loop condition is different in the two code samples. However, only the variables of the operator are switched and both are applying the same comparison.}

  \label{fig:algo_code_sample_sim}
  \vspace{-5pt}
\end{figure}

    %R2-13
     Our observation shows that in some cases however Copilot's suggestions are not exactly the same but they are very similar. R2-25: 
    For example, Figure~\ref{fig:algo_code_sample_sim} shows two code samples for ``Insertion Sort''. The differences between the two code samples are only syntactically in a few lines. Code Sample \#1 calculates the length of the list within the range in for loop instructor. However, Code Sample \#2 assigns the length of the list into a variable and then uses it to control the loop. Also, the comparison operator in the while loop condition is different in the two code samples. However, only the variables of the operator are switched and both are applying the same comparison. 
    
     Therefore, in addition to ``Code Reproducibility'', we report the ``Code Similarity'' as the average similarity between pairs of correct solutions for different fundamental algorithmic problems. To calculate the similarity, we follow the same AST similarity measure as explained above. For ``Within a Trial'', we compare all pairs of correct solutions in different attempts within a trial, and for ``Across Trials'', we compare all pairs of correct solutions between two trials. Finally, the average of these comparisons is reported for each problem.

\subsection{\textbf{RQ2: Copilot vs. Human}}
In this subsection, we aim to describe our research method for RQ2, on how to compare Copilot codes with human written codes in different quantitative metrics. First, we illustrate the dataset of programming tasks that we used in our experiments and explain why we select this dataset. Then, we explain how we employ Copilot to generate solutions for each task in this dataset. After that, we present how we selected students' solutions for this comparison. Finally, we discuss the criteria to compare Copilot with students in solving Python programming tasks from different aspects.

\subsubsection{\textbf{Dataset: Python Programming Tasks}}\label{sec:dataset}
 To address RQ2, as we already discussed in Subsection~\ref{sec:RQ1_all}, we require a dataset of programming problems that Copilot can solve so that we can conduct further investigations on Copilot’s suggestions. Considering the choice of programming tasks and in order to have a fair comparison, we compare Copilot with junior developers. Therefore, we choose a dataset of a Python programming course that includes students' submissions for $5$ programming assignments\footnote{\url{https://github.com/githubhuyang/refactory}}.

There are studies on Copilot that used easy but more practical programming tasks than course assignments, such as tasks in~\cite{vaithilingam2022expectation} (i.e., editing a CSV file), but these types of tasks need less problem-solving effort compared to the assignments of a programming course in our selected dataset.

As in this study, our investigations go beyond the code correctness, there are other advantages to using this dataset. First, this dataset includes students’ submissions that support our research question on comparing Copilot with humans. Second, the task description in this dataset is human-written, reducing the chance of memorization issues \cite{carlini2022quantifying}. They are new tasks for testing Copilot and different from the tasks in the Codex test set, i.e., the HumanEval dataset \cite{chen2021evaluating}. In addition, this dataset includes different test cases for each task alongside a tool for automatically checking the functional correctness of solutions and the repairing cost of buggy solutions.

%\url{https://program-repair.org/benchmarks.html}}

This dataset includes $2442$ ``Correct'' and $1783$ ``Buggy'' student submissions for $5$ Python programming assignments in a Python course. %\Amin{any paper that used this dataset, that we can add?}.
Another study also used this dataset for characterizing the benefit of adaptive feedback for errors generated by novice developers~\cite{ahmed2020characterizing}. Table~\ref{tab:dataset} shows the description of each programming task. Each task includes a description of the problem, one or more reference solutions, a different number of submissions by students that includes ``Correct'' and ``Buggy'' solutions, and different unit tests for each task, with an average of 9 tests per problem, to evaluate the functional correctness of solutions. 

%\Amin{I revised this, please check it!}
%\textbf{\textit{Repairing Tool.}} 
This dataset also contains a tool named ``Refactory'' to automatically repair the buggy submissions of students if applicable~\cite{hu2019re}. In our study, we use this tool to repair buggy solutions generated by Copilot and students to evaluate the complexity of fixing bugs in codes generated by Copilot compared to those of junior programmers. This tool matches each buggy program with the closest correct solution based on its AST structure. Then, it modifies different blocks of the incorrect program to repair its bug(s) and convert it to a correct solution if possible. This tool shows better performance than other state-of-the-art methods in repairing buggy programs such as Clara~\cite{gulwani2018automated}. Despite others that need a large and diverse range of correct solutions, this tool can repair buggy codes even with one or two references (i.e., correct solutions). 
%\Amin{why? to evaluate the complexity/difficulty of fixes, for example?}.

\begin{table*}[ht]
  \centering
  \caption{ \textbf{A summary of the dataset used to compare Copilot with the human in solving simple programming tasks.} The Dataset includes the assignments and submissions of a Python programming course. It includes students' submissions for 5 Python programming tasks~\cite{hu2019re}. The last two columns represent the number of students' submissions in two categories ``Correct'' and ``Buggy''. }
\begin{tabularx}{0.9\textwidth}{ccXcc}
%\begin{tabular}{c|c|c|c|c} 
\specialrule{.1em}{.05em}{.05em}
%& \multirow{-2}{*}{\textbf{Task}} &
& \textbf{Task}&
%\multicolumn{1}{c}{\multirow{-2}{*}{\textbf{Description}}} &
%\shortstack[C]{\textbf{\#Correct}\\\textbf{Solution}} & \shortstack[C]{\textbf{\#Buggy}\\\textbf{Solution}}\\
 \multicolumn{1}{c}{\textbf{Description}}&
\textbf{Correct} &
\textbf{Buggy}
\\
%\midrule
\specialrule{.1em}{.05em}{.05em}
\multirow{4}{*}{q1} & \multirow{4}{*}{\shortstack[C]{\textbf{Sequential}\\\textbf{Search}}} & Takes in a value ``x'' and a sorted sequence ``seq'', and returns the position that ``x'' should go to  such that the sequence remains sorted. Otherwise, return the length of the sequence. & \multirow{4}{*}{768} & \multirow{4}{*}{575}\\
\hline
\multirow{5}{*}{q2} & \multirow{5}{*}{\shortstack[C]{\textbf{Unique Dates}\\\textbf{Months}}} & Given a month and a list of possible birthdays, returns True if there is only one possible birthday with that month and unique day, and False otherwise. Implement $3$ different functions: unique\_day, unique\_month and contains\_unique\_day. & \multirow{5}{*}{291} & \multirow{5}{*}{435}\\
\hline
\multirow{3}{*}{q3} & \multirow{3}{*}{\shortstack[C]{\textbf{Duplicate}\\\textbf{Elimination}}} & Write a function remove\_extras(lst) that takes in a list and returns a new list with all repeated occurrences of any element removed. & \multirow{3}{*}{546} & \multirow{3}{*}{308} \\
\hline
\multirow{6}{*}{q4} & \multirow{6}{*}{\shortstack[C]{\textbf{Sorting}\\\textbf{Tuples}}} &  We represent a person using a tuple (gender, age). Given a list of people, write a function sort\_age that sorts the people and returns a list in an order such that the older people are at the front of the list. You may assume that no two members in the list of people are of the same age.& \multirow{6}{*}{419} & \multirow{6}{*}{357}\\
\hline
\multirow{5}{*}{q5} & \multirow{5}{*}{\shortstack[C]{\textbf{Top\_k}\\\textbf{Elements}}} & Write a function top\_k that accepts a list of integers as the input and returns the greatest k number of values as a list, with its elements sorted in descending order. You may use any sorting algorithm you wish, but you are not allowed to use sort and sorted.& \multirow{5}{*}{418} & \multirow{5}{*}{108}\\
\hline
\textbf{Total} & & & 2442 & 1783 \\

\specialrule{.1em}{.05em}{.05em}
\end{tabularx}
%\end{tabular}
  \label{tab:dataset}
\end{table*}

\subsubsection{\textbf{Solving Programming Problems with Copilot}}\label{subsec:cop_method}
To generate solutions with Copilot, akin to Subsection~\ref{subsec:cop_algo_data}, we feed the description of each programming task in Table~\ref{tab:dataset}, called prompt, to Copilot. At each attempt, Copilot only returns the Top-10 solutions for a prompt. Thus, we do not have access to the rest of the potential suggestions. To inquire about the Copilot's consistency in generating solutions, similar to the previous experiments, we repeat the process. In this setup, we repeat the process $5$ times and each time collect its top $10$ suggested solutions. Expressly, we ask Copilot to solve each programming problem in 5 different attempts and collect the top $10$ suggested solutions in each one. Thus in total, we collect $50$ solutions by Copilot for each problem. 

As we already explained in Subsection~\ref{sec:dataset}, there are different test cases per task. To evaluate the functional correctness of Copilot's solutions, a solution is considered ``'Correct'' if it passes all the unit tests related to its problem. Otherwise, it is considered as ``Buggy''.
%\Amin{what do you mean by category}.

\subsubsection{\textbf{Downsampling Student Solutions}}\label{sec:resample}
In each attempt on Copilot, we only have access to its top $10$ suggestions while the average number of student submissions for these tasks is  $689.8$.  One solution to have more suggestions by Copilot could be to increase the number of attempts on Copilot. But, increasing the number of attempts to more than $5$ will increase the number of duplicate answers in Copilot's suggestions. We discuss the duplicate solutions in Subsection~\ref{sec:criteria} with more details. 

Thus, instead of increasing the number of attempts on Copilot, we downsample the students' submissions to the same size of Copilot solutions (50 in total) to have an equal number of solutions for students and Copilot. 

\subsubsection{\textbf{Evaluation Criteria}}\label{sec:criteria}
For this part of our study, we consider different criteria to compare solutions suggested by Copilot and students to solve these programming tasks. We investigate the solutions on the following markers. In the rest of this section, we explain each metric in more detail.
%\Amin{or metric, Ahura said marker in his part}:
\begin{enumerate}
    \item Correct Ratio (pass@Topk)
    \item Repairing Costs
    \item Diversity
    \item Cyclomatic Complexity
    \item Syntactic Mastery.

\end{enumerate}

%\Amin{I revised, please check out!}\textbf{The Correct Ratio of Solutions.}
\subsubsection*{\textbf{(1) Correct Ratio (pass@Topk)}}
A very common metric to evaluate programming language models is pass@k metric~\cite{li2022competition,chen2021evaluating}. For example, calculating pass@100 shows the fraction of correct solutions out of 100 solutions. %we need to generate $n\geq100$ sample solutions, count the number of solutions that passed all test cases, and then calculate the fraction of $100$ out of all correct solutions. 
However, since Copilot returns only the Top 10 solutions in each attempt, we cannot accurately use this metric in our study.

In this study, what attracts our interest is the pass@Topk in all the attempts. It means that if we call Copilot $n$ times for the same problem (the same prompt), $n$ equals the number of attempts, and collect the Topk solutions of each attempt, then pass@Topk equals the fraction of these solutions that passed all the test units. As an example for pass@Top2, we collect all the Top2 suggested solutions for a problem in $n=5$ different attempts ($\#solutions= k\ast n= 2\ast5= 10$). Then pass@Top2 reports the fraction of these 10 solutions that passed all test units. We can calculate pass@K for Copilot but we cannot calculate it for students.

Another evaluation that comes to our attention is the Correct Ratio (CR) of solutions. We calculate the correct ratio of solutions same as Subsection~\ref{Sec:algo_eval}. Here by CR, we mean the fraction of correct solutions out of all solutions suggested by the Copilot or human for each problem. We calculate this fraction for each problem while collecting Topk suggestions of Copilot in different attempts. For students, we calculate the fraction of correct submissions out of all students' submissions for each problem. 

Also, we calculate the distribution of the CR and its average in independent attempts on Copilot. We like to study how increasing the number of attempts (giving different chances to Copilot to solve the same problem) impacts the CR.

\subsubsection*{\textbf{(2) Repairing Costs}}
After computing the CR for Copilot and students, we aim to compare Copilot's buggy solutions with students' buggy submissions. Our observation shows that several buggy solutions generated by Copilot can be easily converted into a correct solution by applying small changes. We discuss this observation in detail in Subsection~\ref{sec:repair_result}. 

 Repairing cost of bugs in a software project is an important metric to show the quality of a code snippet \cite{kim2006long}. The long repairing time of a bug can be correlated with structural problems in a code snippet \cite{kim2006long}. One of the important factors that impact the repairing time of a bug is the code churn or the size of changes that are required to fix the bug \cite{zhang2012empirical}. Complex or low-quality codes (in case of being buggy) need more time to be repaired, e.g., the developer needs to spend more time to detect the bug, or bigger patches are required for fixing the bug. Thus, to compare the quality of codes generated by Copilot with students, we repair the buggy solutions and then compare them in terms of repair costs.

We use the repairing tool that we explained in Subsection~\ref{sec:dataset}. We choose an automated tool for repairing buggy codes because:
\begin{enumerate}
    \item Using an automated tool to fix bugs is very common in software projects. Software projects train their own tool for automatically fixing the bugs within the projects to save developers time~\cite{arcuri2008automation}.
    \item By using an automated tool, we prevent our repairing process from being biased by one specific human expertise.
\end{enumerate}

This tool reports three different metrics to evaluate the repairing cost of buggy solutions~\cite{hu2019re} including:
\begin{itemize}
    \item \textbf{Repair Rate:} This metric shows the fraction of buggy codes that passed all test cases after the repair process.
    \item \textbf{Avg. Repair Time:} This metric shows the average time taken to repair a buggy program in seconds.
    \item \textbf{Relative Patch Size (RPS):} This metric defines as the Tree-Edit-Distance (TED) between the AST of a buggy code and the AST of its repaired code, normalized by the AST size of the buggy code. 
\end{itemize}

\subsubsection*{\textbf{(3) Diversity}}\label{sec:diversity} 
It is already shown in language models such as Codex that increasing the number of sample solutions for a programming task can increase the number of correct solutions that pass all test units~\cite{chen2021evaluating, li2022competition}. However, they did not study if this increment is due to the increasing diversity of solutions or if the new correct solutions are just a duplication of previous ones.

\begin{figure*}[t]
\centering
\subfloat[Sample code 1]{\includegraphics[width=0.31\textwidth]{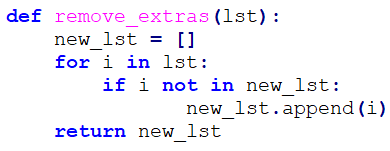}\label{fig:code1}}
  \hfill
  \subfloat[Sample code 2]{\includegraphics[width=0.31\textwidth]{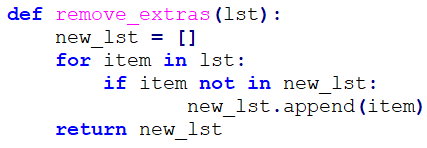}\label{fig:code2}}
  \hfill
  \subfloat[Sample code 3]{\includegraphics[width=0.31\textwidth]{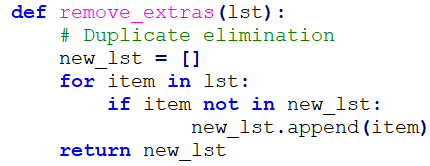}
  \label{fig:code3}}
 \caption{\textbf{ Three different solutions were generated by Copilot for the q3: Duplicate Elimination Task in one attempt.} {There is no difference between the approach of these 3 solutions in solving the task. The only difference between (a) and (b) is in variable names, ``i'' and ``item''. The difference between (c) and (b) is the additional comment in (c). The differences between (c) and (a) are in variable names and comments.}}
  \label{fig:codesamples}
  \vspace{-1em}
\end{figure*}

Copilot claims that it removes duplicate solutions among the Top 10 suggested solutions in a single attempt. However, our observations show the appearance of duplicate solutions in the Top 10 suggestions of a single attempt.  Figure~\ref{fig:codesamples} shows three different solutions generated by Copilot for task q3: Duplicate Elimination at a single attempt. As we can see, the structure of all three codes is the same. The only difference between Figure~\ref{fig:code1} and Figure \ref{fig:code2} is in the variable name, ``item'' and ``i''. Also, the solution in Figure \ref{fig:code3} is the same as the solution in Figure~\ref{fig:code1} alongside comments. Since Copilot compares the sequence of characters to eliminate duplicates, it considers these three solutions as three unique suggestions in the Top 10 solutions of a single attempt.
 
To remove duplicate solutions in each attempt, we use the method discussed in Subection~\ref{Sec:algo_eval} for reproducibility evaluation. We investigate if increasing the number of attempts and consequently increasing the total number of solutions will increase the number of unique solutions. Also, we compare the diversity of solutions (correct and buggy) provided by Copilot and students.  This metric compares Copilot's novelty in generating solutions to that of students in solving a programming task.
  
To remark on the duplicate solutions, as we discussed in Subsection~\ref{Sec:algo_eval}, we compare the AST of two codes. We eliminate the leaves in AST which are related to variable or function names. Also, we ignore comments or any natural language text in each solution. Then, we calculate a similarity between the AST of every two solutions for a problem by the method introduced in~\cite{salazar2020comparing}. If the similarity between two ASTs is equal to 1, then they are assumed to be duplicates. We keep just one of the solutions. Any value less than 1 represents a difference between the functionality of the two solutions.

\subsubsection*{\textbf{(4) Cyclomatic Complexity}}  
A programming language is comprised of a set of programming keywords and built-in functions, methods, and types. Developers may solve a simple programming task in different ways. They may choose different programming keywords and built-ins to solve the same problem. However, even though flexibility in completing a programming task is desired, it can impact the efficiency, readability, and even maintainability of codes in some cases~\cite{maruping2009role, dos2018impacts}. These differences can also reflect developers' mastery of the programming language. For example, Figure~\ref{fig:code_sample_CC} shows two different solutions to a simple programming task, q4: Sorting Tuples, from Table~\ref{tab:dataset}.  Code Sample \#1 has more diverse programming syntax keywords and built-in functions, but Code Sample \#2 is easier to understand and more readable.

%\textcolor{blue}{We investigate the efficiency and readability of Copilot's codes compared to humans with evaluation criteria of Cyclomatic Complexity (C.C.) and Syntactic Mastery explained in Subsection~\ref{sec:criteria}/(5).}

 Cyclomatic Complexity (McCabe's Cyclomatic Complexity C.C.) is another code quality metric that evaluates the understandability of a code snippet. C.C. shows the number of independent paths in a code component, specifically, the number of decisions that can be made in a source code \cite{ebert2016cyclomatic, sarwar2013cyclomatic}. Measuring the understandability of code snippets allows us to estimate the required effort for adding new features to the code or modifying it \cite{scalabrino2019automatically}.

There are studies that apply C.C. to measure the readability and understandability of small code snippets~\cite{fakhoury2019improving,dantas2021readability, NguyenMSR22}. When comparing solutions for a problem, a lower C.C. indicates a more readable and understandable code. For example, in Figure~\ref{fig:code_sample_CC}, the C.C. of code samples \#1 and \#2 are $4.13$ and $1$, respectively. While code sample \#1 represents two nested for-loops to sort the list, code sample \#2 simply calls sort and uses a lambda to loop over the list. Such an approach is more Pythonic and also more understandable.

To evaluate if Copilot’s suggestions are as understandable as humans’, we calculate the C.C. of Copilot’s solutions and compare them to the C.C. of humans’ solutions for the same problems. Thus, we can assess whether Copilot can provide understandable code that is easy to change and maintain (lower C.C.) or not if used as a pair programmer in a software project. We use a Python package, RADON~\footnote{\url{https://radon.readthedocs.io/en/latest}},  to calculate it. C.C. close or above 10 is interpreted as not a best practice code.
%Cyclomatic Complexity (C.C.) (McCabe's Cyclomatic Complexity) shows the number of independent paths. Specifically, the number of decisions that can be made in a source code~\cite{ebert2016cyclomatic, sarwar2013cyclomatic}. 
%\textcolor{blue}{R1-2: When comparing solutions for a problem, the lower C.C. of a code shows the more readable and understandable code. For example, in figure~\ref{fig:code_sample_CC}, the C.C. of sample \#1 and \#2 is $4.13$ and $1$ respectively.} 
\vspace{15pt}
\subsection*{\textbf{(5) Syntactic Mastery}}
 
As we already discussed in Subsection~\ref{sec:criteria}/(4), different syntax patterns and built-in functions, methods, and types in solving the same problem can reflect the developers' mastery as novice developers may not be familiar with all possible programming keywords and features in a programming language. While diversity in syntax patterns of a solution to address a specific task shows familiarity with more programming keywords and built-ins, these diverse solutions may not necessarily be the best practice to solve a problem. One of the goals of pair programming in industrial projects is to transfer such experiences from experts to novice developers~\cite{plonka2015knowledge, lui2006pair, fronza2009interpretation}. So, as another evaluation criterion, we compare the diversity of programming keywords and Python’s built-in functions of Copilot's code to those of humans.

\begin{figure}
\centering
  \includegraphics[width=0.85\linewidth]{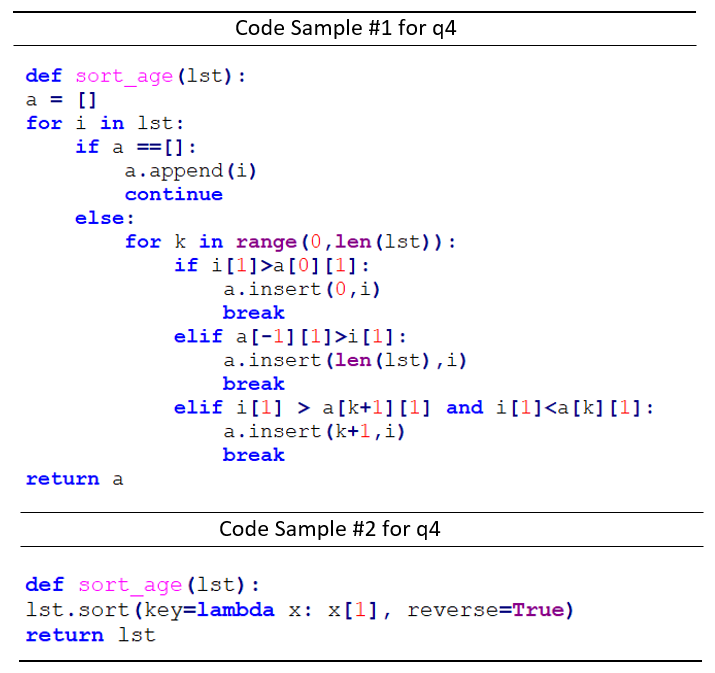}
  \caption{\textbf{Two different solutions to solve q4: Sorting Tuples.} Code Sample \#1 has more diverse syntax patterns and Python built-in functions compared to Code Sample \#2. But \#2 is more readable and less complex (in terms of C.C.) in understanding because of using more advanced programming syntax and built-in methods. The C.C. of Code Sample \#1 (written by a human) is $4.13$ while it is $1$ for Code Sample \#2 (suggested by Copilot).}
  \label{fig:code_sample_CC}
  %\vspace{-1em}
\end{figure}

For example, the codes in Figure~\ref{fig:code_sample_CC} are different solutions to solve the same programming task. Code sample \#1 has more diverse programming syntax keywords such as \{`FunctionDef', `List[None]', `for', `if', `BoolOp', `else', `break', `elif', `return'\} and more diverse built-ins such as \{`append', `range', `insert'\}. Code sample \#2 includes programming syntax keywords such as \{`FunctionDef', `Lambda', `NameConstant', `Subscript[Num]'\} and built-in method, {'sort'}, which are less diverse than the first sample but more advanced and a less complex solution (in terms of cyclomatic complexity).

We follow the instructions suggested by~\cite{moradi2021assessing} to collect programming syntax patterns. We convert each solution to its AST and then walk through the syntax tree to collect nodes as programming keywords.

To collect built-in functions within a code, first, we need to distinguish the built-in function from other function calls since all types of calls in Python, from built-in to local or public library, are a subset of a node named “Call” in AST. To do so, we extract a list of Python built-ins~\footnote{\url{https://docs.python.org/3/library/functions.html}}. Then, we collect the node's name of the node “Call” if its “class\_name” was in the list of Python built-ins. We compare the diversity of the keywords and Python built-ins in Copilot's and humans' codes to study their capabilities in using Python’s keywords and built-ins.

\section{Empirical results}\label{sec:results}
In this section, we present the results we obtained to answer our RQs, one by one.
%compare Copilot with human in solving simple programming problems. We first report our methodology including the dataset we used, evaluation criteria, and experimental design.

\subsection{\textbf{RQ1: Copilot on Algorithm Design}} \label{sec:algorithms}%\Foutse{we can use the RQ as title here so that it clearly reflects the content of the section!}
In this section, we assess the capability of Copilot to solve algorithmic problems. To highlight the difference between our two trials which have been conducted 30 days apart from each other, for each marker, we have indicated the results of the solutions ``Within a Trial'' separately from each other as ``First Trial'' and ``Second Trial''. For this part of our study, we discuss the different evaluation criteria per each category of problems since our finding shows there is a correlation between the difficulty of the categories and the results.

%\subsubsection{\textbf{Response Received}}

  % We tested Copilot on 4 distinct algorithmic problem categories:
    
   % \begin{itemize}
    %    \item \textbf{Sorting algorithms}
     %   \item \textbf{Binary search trees}
      %  \item \textbf{Elementary graph algorithms}
       % \item \textbf{Greedy algorithms}
    %\end{itemize}

    %Tables \ref{table:CAD_SA_results} -- \ref{table:CAD_GA_reproducibility_results} show our results on testing Copilot's ability for code generation. In the following sections, we discuss our results in detail. %\Foutse{we discuss each result in details!}.
    %To highlight the difference between our 2 trials which have been conducted 30 days apart from each other, for each marker, we have indicated the results of the trials separately from each other. To showcase our results from each round of testing in each trial, we count the number of successful tests. For example, if the generated code passes 2 of the 3 tests for a given marker, we display it as ``2/3''. We have also used ``0'' and ``-'' to differentiate between cases where we obtained no results and not applicable cases. %, respectively. 
    %For example, if we did not receive a response for a solution, we display it as ``0/3'' and since we had no responses, finding a correct solution was not applicable, hence, we use ``-'' to indicate this circumstance.

    \subsubsection{\textbf{Sorting Algorithms}}
     In this section, we discuss our findings on Sorting Algorithms. For those evaluation metrics where the manual inspection of authors was required (Response Received, algorithm validation on Correct Ratio, and code Optimality), the authors achieved 89\% of the Kappa agreement. We discuss the details in the following of this section. 
    
    \subsubsection*{\textbf{(1) Response Received}}
    Our results in Table~\ref{table:CAD_SA_results} on sorting algorithms show that when the algorithm gets more difficult and requires more details in implementation, Copilot struggles to generate solutions. For example, on the first trial, for Heap and Radix sort, Copilot generates code in only one of the 3 attempts within the trial. However, in the second trial, Copilot showed improvement as it generated codes in all three attempts. The situation is the opposite for Merge sort. In the first trial, Copilot generates code in all three attempts. But in the second trial, it is responsive in only two of the three attempts.

  \subsubsection*{\textbf{(2) Correct Ratio}}
  Copilot shows various behavior in generating correct solutions for sorting algorithms. The difficulty of problems impacts its ability to generate a correct solution and to use the correct algorithm for the implementation. However, Copilot shows different behavior in two different trials. For example in the first trial for bubble and bucket sort which are two easy sorting algorithms, 100\%, and 85.71\% of Copilot's suggestions were correct respectively. However, in the second trial, it generates no correct solutions for these two sorting problems.
  
  Since implementing heap sort requires implementing a max heap, and then writing a sorting function, this algorithm is harder to implement. In the first trial, Copilot generates no correct solution for this problem. However, during our second trial, 9.09\% of its suggestions for this problem are correct. In the second trial for Radix sort, Copilot showed improvement in solving the problem as it generated codes in all three attempts (Response Received) but none of the generated codes were correct.

   Copilot shows some particular behavior for some of the sorting algorithms. For example, during the second trial where we asked it to generate codes for Bucket sort, some of the generated codes were calling the Quick sort function for sorting the buckets even though Quick sort had not been implemented in the code.
  
  For validating the algorithm choice in solutions that passed all unit tests, two authors disagreed on the result for selection sort. The input prompt was summarized from the descriptions collected from the algorithm design book~\cite{cormen2022introduction}. The given prompt for this algorithm was ``Create a function that accepts a list as input. The function should create two lists named sorted and unsorted. The function sorts an array by repeatedly finding the minimum element (considering ascending order) from an unsorted list and putting it at the beginning of the sorted list. Finally, it should return the sorted array''. Given this description, the second author only accepted solutions that followed this exact description, mainly those which created the two empty \textit{sorted} and \textit{unsorted} lists. Upon review, however, the first and third authors mentioned that some solutions followed the selection sort algorithm, without following the exact steps mentioned in the description. After discussions, these solutions were considered as correct as well.

   \subsubsection*{\textbf{(3) Code Optimality}}
    Our result on code optimality shows that if Copilot is able to generate correct solutions for a sorting algorithm, it generates an optimal solution for that problem, too. In the first trial, Copilot generates correct solutions for 7 out of 8 sorting algorithms and it is able to generate optimal solutions for these 7 problems in the same trial, too. In the second trial, Copilot generates correct solutions for 4 sorting algorithms and it is able to generate optimal solutions within its correct solutions for these 4 sorting problems as well.
   
    Since we have no correct solutions for example for Bubble sort or Bucket sort in the second trial, code optimality is not applicable for these cases. Thus, we show their results with ``-''.

    However, this result on sorting algorithms is due to the fact that for some of the sorting algorithms such as bubble sort or heap sort, there is no other possible implementation than the optimal one (quadratic or log-linear).

\begin{figure}
\centering
  \includegraphics[width=0.85\linewidth]{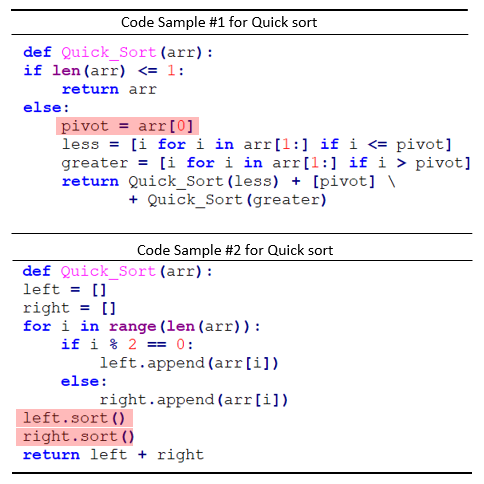}
  \caption{\textbf{ Two different solutions suggested by Copilot for Quick sort.} Code Sample \#1 is a recursive function. It picked the first element as a pivot to partition the given array and employed the correct ``Divide and Conquer'' algorithm to implement Quick sort. However, Code Sample \#2 randomly divided the given array into partitions. It is buggy, and it is not deploying the sorting properly, but it uses the Python built-in function, ``sort'' to sort each partition.}

  \label{fig:quick_sort}
  \vspace{-5pt}
\end{figure}

     We also observed that Copilot generates Pythonic code or uses Python's language-specific features instead of re-implementing the desired functionality to some extent. For example, alongside using list comprehensions (which are faster in Python than iterating over the list in explicit for-loops), Copilot generated codes that use built-in functions. An example of this can be observed in Figure~\ref{fig:quick_sort}, codes generated for the Quick sort where after dividing the input array into left and right subarrays, instead of generating code for sorting the arrays, Copilot used Python's built-in sort function. For sorting problems where iterating over the entire input was required, instead of using while loops, Copilot generates code with either explicit “for” loops or list comprehensions. Doing so removes the risk of getting trapped in an infinite loop and in the case of using list comprehensions can make a real difference in the program’s running time.
    
   %\textcolor{blue}{Since we have no correct solutions for example for Bubble sort or Bucket sort in the second trial, code optimality is not applicable for these cases. Thus, we show their results with ``-''.}

    %\textcolor{blue}{However, this result on sorting algorithms is due to the fact that for some of the sorting algorithms such as bubble sort or heap sort, there is no other possible implementation than the optimal one (quadratic or log\-linear).}

    %\textcolor{blue}{R1-23 and R1-10: In some cases, we observed that Copilot generates Pythonic (code that uses Python's features instead of re-implementing them) to some extent. For example, alongside using list comprehensions (which are faster in Python than iterating over the list in explicit for loops), Copilot generated code that used built-in functions. An example of this can be observed in codes generated for Quick sort where after dividing the input array into left and right components, instead of generating code for sorting them, Copilot called the built-in sort function to sort them. For sorting problems where iterating over the entire input was required, instead of using while loops, Copilot generates code with either explicit “for” loops or list comprehensions. Doing so removes the risk of getting trapped in an infinite loop and in the case of using list comprehensions can make a real difference in the program’s running time.}

    \subsubsection*{\textbf{(4) Code Reproducibility and Similarity}}
     As the last evaluation metric in Table~\ref{table:CAD_SA_results}, we report if at least one of the correct solutions suggested by Copilot is exactly reproduced (similarity equals 1) within a trial and across two trials. Our result shows that correct solutions are not exactly reproduced for the majority of sorting algorithms.

    However, the correct solutions are not exactly reproduced within a trial or across two trials, our results in Table~\ref{table:CAD_SA_reproducibility_results} on the similarity degree between pairs of correct solutions show that they are very similar in some cases. For example, for Quick sort in the second trial, the correct solutions are not exactly reproduced but based on Table 3, there is a 0.99 similarity between its correct solutions. As another example, for selection sort in the first trial and across two trials, the correct solutions are not exactly reproduced but the similarity degree between them equals 0.61 and 0.63 respectively.

    Same as code optimality, if Copilot was not able to generate the correct solution for a problem within a trial, then the code reproducibility metric and similarity degree are not applicable. Also, if Copilot generates correct solutions for the sorting problems just in one of two trials, then reproducibility and similarity across trials are not applicable. We use ``-'' for nonapplicable cases.
     \paragraph{\textbf{Summary of Results}}
     In summary, Copilot is relatively capable of providing solutions for sorting problems. It is responsive (Response Received) for 6 out of 8 sorting problems of the first trial and for 7 out of 8 problems of the second trial.
     
     On Correct Ratio, in the first trial, Copilot generates correct solutions for all sorting problems except Heap sort, and on average, 51.42\% of its solutions within this trial are correct. However, in the second trial, it generates correct solutions for only 4 sorting problems (out of 8) and the average correct ratio within this trial is 34.77\%. In both trials, if Copilot generates the correct solution for a problem, at least one of those correct solutions is optimal.
     
     Finally, the correct solutions suggested by Copilot are not exactly reproduced for the 4th, 3rd, and 2nd sorting problems in the first trial, second trial, and across two trials respectively, but the similarity degree between some of those non-reproduced correct solutions is above 0.6. In some trials and for some of the sorting problems the code optimality, reproducibility, and similarity are not applicable due to the lack of comparable correct solutions.
     
            \begin{table*}
            \centering
            %\large
            \caption{\textbf{ Results of Copilot's code generation ability on fundamental algorithmic problems.} \textbf{``Response Received''} shows the number of attempts in each trial that Copilot can generate codes for the proposed prompt. It ranges in $[0,3] \in N $. \textbf{``Correct Ratio''} shows the percentage of correct solutions in each trial. \textbf{``Optimal''} status is ``Yes'' if at least one of the correct solutions in each trail is optimal. If at least one correct solution in one of the three attempts (Within a Trial) repeats in two other attempts (at the same Trial), then \textbf{``Reproduced''} is ``Yes''. \textbf{``Across Trials''} for \textbf{``Reproduced''} metric is ``Yes'' if at least one of the correct solutions from the First Trial repeats in Second Trial.
            If the metrics are not applicable then it presents by ``-''. For example, in ``Second Trial'' of ``Bubble Sort'', we receive no (0) response from Copilot. Consequently, ``Correct Ratio'' and ``Optimal'' in `Second Trial'' and ``Reproduced'' in ``Second Trial'' and ``Across Trials'' assign ``-'' for this algorithm.}
           \resizebox{0.95\textwidth}{!}{
            \begin{tabular}{l c c c c c c c c c}
            \toprule

  \multirow{2}{*}{\textbf{Algorithm}}
                & \multicolumn{2}{c}{\textbf{Response Received [0,3]}}
                & \multicolumn{2}{c}{\textbf{Correct Ratio [\%]}}
                & \multicolumn{2}{c}{\textbf{Optimal [Yes/No]}}
                & \multicolumn{3}{c}{\textbf{Reproduced [Yes/No]}}\\
               
                & \textbf{First Trial}&\textbf{Second Trial} 
                & \textbf{First Trial}&\textbf{Second Trial} 
                & \textbf{First Trial}&\textbf{Second Trial}
                & \textbf{First Trial}&\textbf{Second Trial}&\textbf{Across Trials}\\ 
                 
            \midrule
            \multicolumn{10}{c }{\textbf{Sorting Algorithms}}\\
  \cmidrule(lr){1-10}
                \textbf{Bubble Sort} & 3 & 3 & 100 & 0 & Yes & - & Yes & - & - \\
                \addlinespace
                
                \textbf{Bucket Sort} & 3 & 3 & 85.71 & 0  & Yes & - & Yes & - & -   \\
                \addlinespace
                
                \textbf{Heap Sort} & 1 & 3 & 0 & 9.09 & - & Yes & - & No&- \\
                \addlinespace
            
                \textbf{Insertion Sort} & 3  & 3 & 100 & 100 & Yes & Yes & Yes & Yes & Yes\\
                \addlinespace
                
                \textbf{Merge Sort} & 3 & 2 & 33.34 & 0& Yes & - & No & - & -\\
                \addlinespace
                
                \textbf{Quick Sort} & 3  & 3 & 16.67 & 16.67& Yes & Yes & No & No &No \\
                \addlinespace
                
                \textbf{Radix Sort} & 1  & 3 & 10 & 0 & Yes & - & No & -& - \\
                \addlinespace
                
                \textbf{Selection Sort} & 3 & 3  & 14.28 & 13.34& Yes & Yes & No & No&No \\
                
 % \midrule
 % \midrule
  \cmidrule(lr){1-10}
                 \multicolumn{10}{c }{\textbf{Binary Search Trees}}\\
    \cmidrule(lr){1-10}
     \textbf{Data Structure} & 3 & 1 & 61.9 &	35.71
 & Yes & Yes & No & No & Yes \\
                \addlinespace
                 \textbf{Min and Max Values in Tree} & \centering 3 & 3 & 71.42&	66.67
 & Yes & Yes & No & Yes& No \\
                \addlinespace
                
                \textbf{In-order Tree Walk} & 3 & 3 & 94.12&	16.67
 & Yes & Yes & Yes & No& Yes\\
                \addlinespace
            
                \textbf{Finding The Successor Node} & 3  & 3 & 100 & 100 & No & Yes & No & Yes& Yes\\
                \addlinespace
  \cmidrule(lr){1-10}
                 \multicolumn{10}{c }{\textbf{Elementary Graph Algorithms}}\\
\cmidrule(lr){1-10}
                 \textbf{Simple Data Structure} & 2 & 2 & 50	& 0 & Yes & - & No & - & -\\
                \addlinespace
                
                \textbf{Breadth First Search} & \centering 3 & 3 & 100&	100 & Yes & Yes & Yes & Yes&Yes\\
                \addlinespace
                
                \textbf{Depth First Search} & \centering 3 & 3 & 75&	0 & Yes & - & No & -&-\\
                \addlinespace
                
                \textbf{Directed Acyclic Data Structure} & 2  & 3 & 86.37 &	0 & Yes & - & No & - & -\\
                \addlinespace
                
                \textbf{Finding Reachable Vertices} & 3  & 3 & 60 &	100 & Yes & No & Yes & Yes& No\\
  \cmidrule(lr){1-10}
 
                 \multicolumn{10}{c }{\textbf{Greedy Algorithms}}\\
  \cmidrule(lr){1-10}
  
                \textbf{Activity Class} & 2 & 3 & 0 & 0 & - & - & - & -& - \\
                \addlinespace
                
                \textbf{Comparing Activities} & \centering 3 & 3 & 9.52 & 0 & Yes & - & Yes & - &- \\
                \addlinespace
                
                \textbf{Adding Activities to a Set} & 3  & 3 & 13.33 & 16.67 & Yes & No & No & No & Yes\\
                \addlinespace
                
                \textbf{Generate All in one Prompt} & 1  & 3 & 0 &  0 & - & - & - & -&-\\

            \bottomrule
            \end{tabular}}
        \label{table:CAD_SA_results}
        \end{table*}

\begin{table}[htbp]
            \centering
            \caption{\textbf{ Similarity ratios of the AST of Copilot’s correct suggestions on fundamental algorithmic problems.} To calculate the similarity, we removed the duplicate correct solutions in each attempt (three attempts within a trial). The results show however some of the correct solutions are not exactly reproduced in different attempts within a trial or between two trials, but they are very similar. The similarity is blank, ``-'', if it cannot be calculated (i.e. no correct solution or only one correct solution). }
           \small
           \resizebox{0.95\linewidth}{!}{
            \begin{tabular} {l c c c}
            %{p{4cm} p{3cm} p{3cm} p{2cm}}
            \toprule
                \multirow{1}{*}{\textbf{Algorithm}}

                & \textbf{First Trial} &
                \textbf{Second Trial} &
                \textbf{Across Trials}\\
                
            \midrule
            \multicolumn{4}{c }{\textbf{Sorting Algorithms}}\\
  \cmidrule(lr){1-4}
                \textbf{Bubble Sort} & 0.93 & - & -\\
                \textbf{Bucket Sort} & 1 & - & - \\
                \textbf{Heap Sort} & - & - & -\\
                \textbf{Insertion Sort} & 0.99 & 1 & 0.99\\
                \textbf{Merge Sort} & - & - & -\\
                \textbf{Quick Sort} & - & - & 0.99\\
                \textbf{Radix Sort} & - & - & -\\
                \textbf{Selection Sort} & 0.61 & - & 0.63\\
                \cmidrule(lr){1-4}
                 \multicolumn{4}{c }{\textbf{Binary Search Trees}}\\
    \cmidrule(lr){1-4}
    \textbf{Data Structure} & 0.51 & 0.46 & 0.53\\
                \textbf{Min and Max Values in Tree} & - & 1 & 0.83 \\
                \textbf{In-order Tree Walk} & 1 & - & 1\\
                \textbf{Finding The Successor Node} & 0.33 & 0.99 & 0.55\\
     \cmidrule(lr){1-4}
                 \multicolumn{4}{c }{\textbf{Elementary Graph Algorithms}}\\
\cmidrule(lr){1-4}
 \textbf{Simple Data Structure} & 0.25 & - & -\\
                \textbf{Breadth First Search} & 0.54 & 0.72 & 0.45 \\
                \textbf{Depth First Search} & 0.73 & - & -\\
                \textbf{Directed Acyclic Data Structure} & 0.63 & - & -\\
                \textbf{Finding Reachable Vertices} & 0.79 & 1 & 0.076\\

  \cmidrule(lr){1-4}
                 \multicolumn{4}{c }{\textbf{Greedy Algorithms}}\\
  \cmidrule(lr){1-4}
  \textbf{Activity Class} & - & - & -\\
                \textbf{Comparing Activities} & 1 & - & - \\
                \textbf{Adding Activities to a Set} & 0.09 & 0.11 & 0.17\\
                \textbf{Generate All in one Prompt} & - & - & -\\
    
            \bottomrule
            \end{tabular}}
        \label{table:CAD_SA_reproducibility_results}
        \end{table}

        %The authors disagreed on the result of generated codes for selection sort. Our source for describing the problems was \cite{cormen2022introduction} and therefore, the input prompt was summarized from the description in the book. The given prompt for this algorithm was ``create a function that accepts a list as input. the function should create two lists named sorted and unsorted. the function sorts an array by repeatedly finding the minimum element (considering ascending order) from unsorted list and putting it at the beginning of the sorted list. finally it should return the sorted array''. Given this description, the second author only accepted solutions that followed this exact description, mainly those which created the two empty \textit{sorted} and \textit{unsorted} lists. Upon review however, the first and third authors mentioned that some solutions followed the selection sort algorithm, which can be found on the web, without following the exact steps mentioned in the description. After discussions, these solutions were considered as correct as well.
        
    \subsubsection{\textbf{Binary Search Trees}}
         In this section we discuss our findings on Binary Search Trees (BSTs). The Kappa agreement between two authors on evaluation metrics that needed manual inspection  is 100\%.
         For this problem, we first asked Copilot to generate the BST data structure which should comprise of a class with parent, right, and left nodes alongside the node's value. After that, we asked Copilot to generate a method which handles insertion per the BST insertion algorithm for the class. Then, we asked Copilot to create a method for deleting a node. These operations require the BST to be re-built in order to conform to the BST property. We also asked Copilot to implement a search method for finding if a value is present in the BST. These 3 methods, comprise the base BST data structure. In the next steps, we asked Copilot to generate functions for finding the maximum and minimum value in a tree, performing an in-order tree walk, and finding the successor node of a child node. We discuss the details of our results in the following section.
        
        \subsubsection*{\textbf{(1) Response Received}}
         Our results show that Copilot is capable of understanding the BST problems in both trials. Only in the second trial, Copilot struggles in suggesting code in 2 out of 3 attempts for generating the data structure of a BST.
        
        \subsubsection*{\textbf{(2) Correct Ratio}}
        Our results in Table~\ref{table:CAD_SA_results} show that Copilot has inconsistent behavior in generating correct solutions for some BST problems in two trials. For example, considering the ``In-order Tree Walk'',  94.12\% of Copilot's suggestions are correct in the first trial, but in the second trial, it reduces to 16.67\%. However, for the two problems, ``Min and Max Values in Tree'' and ``Finding The Successor Node'', the correct ratio on both trials are very close to each other. For example, for `Finding The Successor Node'', 100\% of Copilot's suggestions are correct in both trials. 
        
        \subsubsection*{\textbf{(3) Code Optimality}}
        It should be noted that, in a majority of the cases, Copilot was able to generate code consistent with optimal time complexities as required for an efficient BST problem. In addition, Copilot was able to generate multiple different versions (with iterative and recursive programming techniques) for ``Finding maximum and minimum values in the tree'', ``In-order tree walk'', and ``Finding successor nodes'' problems. For the ``In-order Tree Walk'' problem, Copilot generated functions inside the main function responsible for executing the walk. These functions were duplicate functions of those generated for finding minimum and maximum values in the tree. This is bad programming practice as it over-complicates the code. However, since these functions were not used by the original function at all, the generated code was still optimal. Copilot tends to generate recursive functions when the solution can be solved using such an approach. For example, for the ``In-order Tree Walk'' and ``Finding maximum and minimum values in the tree'' problems, the generated codes are all recursive functions.
        
        Thus, for all the BST problems in both trials, except for ``Finding successor nodes'' in the first trial, at least one of the correct solutions suggested by Copilot has optimal time complexity.
        
        \subsubsection*{\textbf{(4) Code Reproducibility and Similarity}}
         As it is shown in Table \ref{table:CAD_SA_results}, in the first trial, Copilot exactly reproduces at least one of its correct solutions in 3 different attempts only for the ``In-order tree walk'' problem. Based on Table~\ref{table:CAD_SA_reproducibility_results}, the similarity between the pairs of its correct solutions is not greater than 0.51 for those correct solutions that are not exactly reproduced. For example, the similarity of correct solutions in different attempts of the first trial for ``Data Structure'' and ``Finding successor nodes'' are 0.51 and 0.33 respectively. 
        
        In the second trial, based on Table~\ref{table:CAD_SA_results}, the exact correct solutions are reproduced for ``Finding maximum and minimum values in the Tree'' and ``Finding successor nodes'' problems. The similarity for correct solutions of `Data Structure'' which is not exactly reproduced in this trial is 0.46.
        
       Unlike sorting algorithms, reproducibility across two trials was not an issue on BST problems as Copilot reproduces at least one of the correct solutions from the first trial in the second trials for all BST problems except ``Finding maximum and minimum values in the Tree''. However, Table~\ref{table:CAD_SA_reproducibility_results} shows that the similarity of the correct solution for this problem across two trials is 0.83.
        
%As our results show, Copilot is adept at producing algorithms for BSTs.  As Table \ref{table:CAD_BST_results} shows, unlike sorting algorithms, reproducibility was not an issue as Copilot generated the same code at least once throughout different experiments during different trials. However, as Table \ref{table:CAD_BST_reproducibility_results} shows, there are differences between the codes that are suggested during each experiment which means that Copilot was not simply repeating itself.

        \paragraph{\textbf{Summary of Results}}
        In summary, Copilot is capable of understanding the description of BST problems in both trials, except for the ``Data Structure'' problem on the second trial.

         Copilot has inconsistent behavior in generating correct solutions in two trials as 81.86\% of its solutions are correct in the first trial but the correct ratio equals 54.76\% in the second trial. Copilot was able to generate optimal code for all the BST problems in both trials except for ``Finding successor nodes'' in the first trial. 
         
         Copilot struggled in exactly reproducing its correct solutions within each trial and the similarity of those solutions that are not exactly reproduced is not above 0.51. However, Copilot reproduces at least one of its correct solutions from the first trial in the second trial (Across Trials) for all BST problems except ``Finding maximum and minimum values in the Tree''. Although the correct solutions for this problem are not exactly reproduced across two trials, the similarity of its correct solutions is 0.83.

     %\Foutse{hence? what is the implication of the result?}
    \subsubsection{\textbf{Elementary Graph Algorithms}}

        In this section, we discuss our findings on Elementary Graph Algorithms. The Kappa agreement between the two authors on metrics that needed manual inspection  is 83\%. As our algorithms are becoming more complex, it is required for Copilot to generate code that uses the previous codes that it has generated.  We discuss the details of our results in the following section.
        
        \subsubsection*{\textbf{(1) Response Received}}
         Our results in Table~\ref{table:CAD_SA_results} show that like BSTs, Copilot is adept at generating code for elementary graph algorithms. In the first trial, Copilot generates code in all 3 attempts for all graph problems except ``Simple Data Structure'' and ``Directed Acyclic Data Structure'' and in the second trial, it struggles only in one of the 3 attempts on ``Simple Data Structure''. 
        
        \subsubsection*{\textbf{(2) Correct Ratio}}
        As we can find in Table~\ref{table:CAD_SA_results}, same as BST problems, Copilot shows inconsistent behavior in generating correct
solutions for some graph problems in two trials. For example, for ``Simple Data Structure'', ``Depth First Search'' and ``Directed Acyclic Data Structure'' in the first trial, 50\%, 75\% and 86.37\% of Copilot's Suggestions are correct respectively. However, in the second trial, Copilot is not able to generate correct solutions for these problems. For the ``BFS'' problem, 100\% of Copilot solutions are correct in both trials.

Our observation shows that during different attempts on Copilot to generate code for BFS and DFS, Copilot generated code for both algorithms regardless of us asking it to do so only for one of them.

Even though Copilot was able to recognize and generate code for our description, some of the generated codes had one flaw and since successor methods use the previous methods, this bug was present in every piece of generated code. This snow-balling effect has affected our Kappa score as well. This bug was a result of Copilot considering the nodes being named by integer numbers. As a result, if a node is created with a name that is not an integer (e.g. "A" or "Node1" instead of "1" or "2"), the code will fail to iterate through the list of nodes and generate a syntax error. However, since the code functioned correctly given the normal usage, we labeled them as correct. 

        \subsubsection*{\textbf{(3) Code Optimality}} 
         In the first trial, Copilot generated one optimal solution for each of the graph problems. However, in the second trial, out of 2 problems that Copilot addressed correctly, only one of them, BFS, includes the optimal solution within its correct solutions.
    Checking if a graph is cyclic, requires using a BFS or DFS approach. If Copilot does not use the codes that it has generated for BFS and DFS during checking if a graph is cyclic, we will be left with code pieces that repeat the same operation over and over which is a bad practice in programming. We consider those suggestions as non-optimal.

\begin{figure}
%\centering
  \includegraphics[width=0.6\linewidth]{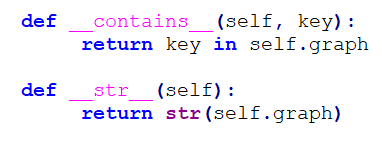}
  \caption{\textbf{R2-10 and R2-25: Code sample of operator overloading.} ``operator overloading'' is an advanced Python feature in which a Python built-in function is re-written by the programmer to behave differently depending on the arguments that it receives as input. ``contains'' and ``str'' are two Python native functions that Copilot re-wrote in graph problems.}
  \label{fig:op_overload}
  \vspace{-5pt}
\end{figure}

 We examined the solutions suggested by Copilot for constructing the graph data structure and observed that its solutions contain both list comprehensions and explicit “for” loops. In one of the correct solutions, the generated code constructs the nodes from the input using explicit “for” loops and in another solution, it does so using list comprehensions. We accept the code that uses list comprehensions as optimal since if the input is large, there is a real running time difference between these two approaches.
 We also observed that some of the generated codes are using an advanced Python feature called ``operator overloading'' in which a native Python function is re-written by the programmer to behave differently depending on the arguments that it receives as input. Figure~\ref{fig:op_overload} shows an example of operator overloading generated by Copilot.
%\Arghavan{here I put the text from backup version as comment in overleaf}
%\textcolor{blue}{R-1-11: We observed that some of the generated code is using an advanced Python feature called ``operator overloading'' in which a native Python function is re-written by the programmer to behave differently depending on the arguments that it receives as input. Figure~\ref{fig:op_overload} shows an example of operator overloading generated by Copilot. Also, we observed Copilot's behavior in using list comprehensions against explicit for loops in constructing the graph data structure. In one of the correct solutions, the generated code constructs the nodes from the input using explicit for loops and in another it does so using list comprehensions. We accept the code that uses list comprehensions as optimal since if the input is large, there is a real speed difference between these two approaches.}

        \subsubsection*{\textbf{(4) Code Reproducibility and Similarity}}
         As we can find in Table~\ref{table:CAD_SA_results}, in the first trial, Copilot is able to reproduce at least one of its correct solutions for only two graph problems, ``Breath First Search'' and ``Finding Reachable Vertices''. However, for other problems such as ``Depth First Search'' and 'Directed Acyclic Data Structure'', the correct solution is not exactly reproduced by Copilot but their similarity equals 0.73 and 0.63 respectively. In the second trial, Copilot is able to reproduce the correct solutions for those two problems that it addressed correctly. For across trials, Copilot is able to exactly reproduce the correct solutions only for the BFS problem.  The similarity between correct solutions of ``Finding Reachable Vertices'' is very low across two trials, 0.076.   
        \paragraph{\textbf{Summary of Results}}
        Our results show Copilot is adept at generating code for elementary graph algorithms. However, same as BST, Copilot shows inconsistent behavior in generating correct solutions for some graph problems in two trials. In the first trial, Copilot is able to generate correct solutions for all graph problems with an average correct ratio of 74.27\%. However, in the second trial, it is able to generate correct solutions for only two problems and 100\% of its correct solutions are correct. Copilot was able to generate optimal code for all problems that it addressed correctly in both trials except for ``Finding Reachable Vertices'' in the second trial. In the manner of reproducibility, it struggled to reproduce its correct solutions for all graph problems. However, the similarity between correct solutions for some problems is more than 0.6.

        %Our results show that like BSTs, Copilot is adept at generating code for elementary graph algorithms. As mentioned, up until now, we described the problem to Copilot in detail without mentioning the name of what we wanted explicitly. However, now that we are asking Copilot the name of the algorithm, the generated codes have little to no differences with each other even through different sets of trials. We observed that some of the generated code is using an advanced Python feature called ``operator overloading'' in which a native Python function is re-written by the programmer to behave differently depending on the arguments that it receives as input. During our testing for BFS and DFS, Copilot generated code for both algorithms regardless of us asking it to do so only for one of them.

    \subsubsection{\textbf{Greedy Algorithms}}
    \label{subsec:ga}
        In this section, we discuss our findings on the ``activity selection'' problem as a Greedy Algorithm. The Kappa agreement between the two authors on metrics that needed manual inspection  is 100\%.  
       The ``activity selection'' problem requires the programmer to define a class for ``activities''. Each activity has a start and end time. The goal of this problem is: given a set of activities where each activity has its own start and ending time, return a set that contains the maximum number of activities that can be performed as long as they do not overlap. Overlapping is defined as:
        
        \begin{itemize}
            \item An activity's start time must be after a previous activity's end time.
            \item An activity should not happen during another activity.
        \end{itemize}
        
         For this problem, we asked Copilot to generate codes for implementing the activity class, comparing activities, and finally checking for overlaps between activities to investigate if the generated solutions are ``greedy''.
        
        \subsubsection*{\textbf{(1) Response Received}}
         Our results in Table~\ref{table:CAD_SA_results} show that Copilot is capable of understanding what the underlying problem is and can generate code for it. Our observations show that Copilot can even generate code when we give it the entire problem definition (activity class, comparing activities, and adding activities to a set) in one go.
        
        \subsubsection*{\textbf{(2) Correct Ratio}}
        Even though Copilot is capable of understanding what we ask from it, the codes that it generates for solving the problem are either buggy or incorrect. For example, given the prompt ``implement a class called activity. Each instance of this class has two attributes: start-time and end-time. Both should be integer numbers between 0 and 24'', the generated code has no functionalities for checking the input type or their boundaries. In another problem, when we asked Copilot to implement a method for comparing activities, we gave it the following prompt: ``implement a function for comparing two activities. the function should return True if the first activity ends before the second activity starts. if the inputs have overlapping start-times, return False''. Here, Copilot implemented the description correctly. However, since this method is dependent on its inputs being instances of the activity class, this code will fail if the input is anything else. Type checking is important and a basic operation to do which Copilot fails to do here. Finally, for adding activities to a set of activities, Copilot was asked to create a method which accepts a set of activities alongside a start time and end time of an activity. The method should first create a new activity instance with the given start and end time and then check if this new activity does not overlap with the activities in the set. Copilot was unable to generate the necessary code for this no matter how detailed the description was.
        
        \subsubsection*{\textbf{(3) Code Optimality}}
        As Copilot was not able to generate correct solutions to most of the problems, we could only analyze the optimality of the solutions generated for ``Comparing activities'' and ``Adding Activities to a Set''. Here, the generated codes were simple (As was the underlying problem) and the solutions required only checking the boundaries of class attributes or whether the output of a function was true or not.
        
        \subsubsection*{\textbf{(4) Code Reproducibility and Similarity}}
        As Table~\ref{table:CAD_SA_results} and ~\ref{table:CAD_SA_reproducibility_results} show, Copilot was only capable of reproducing solutions to a problem for the ``Adding activities to a set'' problem across trails and these solutions were different from each other. As Table ~\ref{table:CAD_SA_reproducibility_results} shows, for the ``Comparing Activities'' problem, Copilot generated solutions which were exactly the same in the same trial. However, in the second trial it was not capable of even producing a correct solution.
        
        \paragraph{\textbf{Summary of Results}}
        The activity selection problem was used as a proxy to see whether Copilot would be able to generate code for solving this problem with a greedy solution. However, Copilot was not able to generate solutions that satisfied the criteria of a correct solution. In particular, Copilot showed difficulties in understanding type checking and variable boundary checking even though such behaviors were explicitly required in the prompt.

\begin{tcolorbox}[colback=blue!5,colframe=blue!40!black]
\textbf{Findings:} Copilot is able to recognize fundamental algorithms by their names and generate correct, optimal code for them as long as the descriptions are short and concise. In some cases, the developers may need to invoke Copilot multiple times in order to receive solutions that are correct and tailored to their descriptions.\newline 
\textbf{Challenges:} Copilot is unable to generate code for type-checking variables. It also generates needlessly complicated code for some simple descriptions. Hence, Copilot still needs to be improved to truly be considered as a pair programmer.
\end{tcolorbox}
        
\subsection{\textbf{RQ2: Copilot vs. Human in Solving Programming Problems}}\label{sec:humans}
%\Amin{For Arghavan!}
%Then, we report our findings. 
%%%%%%%%%%%%%%%%%%%%%%%%%%%%%%%%%%%%%%%%%%%%%%%%%%%%%%%%%%%%%%%%%%%%%%%%%%%%%%%%%%%%%%%%%%%%%%
%%%%%%%%%%%%%%%%%%%%%%%%%%%%%%%%%%%%%%%%%%%%%%%%%%%%%%%%%%%%%%%%%%%%%%%%%%%%%%%
In this section, we discuss our findings to answer RQ2. We discuss the results for each criterion of our evaluation separately.

\subsubsection{\textbf{Correct ratio of Copilot's suggestions and students' submissions}}
As explained in Subsection~\ref{sec:criteria}, we calculate the pass@Topk for solutions generated by Copilot for each programming task. The pass@Topk shows the fraction of correct solutions among the Topk solutions, collected from $5$ different attempts. We normalized the values to report this metric for the programming tasks. 

\begin{figure*}[t]
\centering
\subfloat[Normalized pass@Topk of 5 different attempts]{\includegraphics[width=0.5\textwidth]{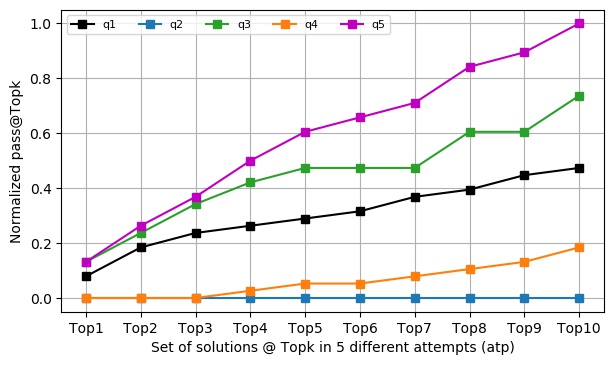}\label{fig:top_cor}}
  \hfill
  \subfloat[CR of solutions in 5 attempts]{\includegraphics[width=0.5\textwidth]{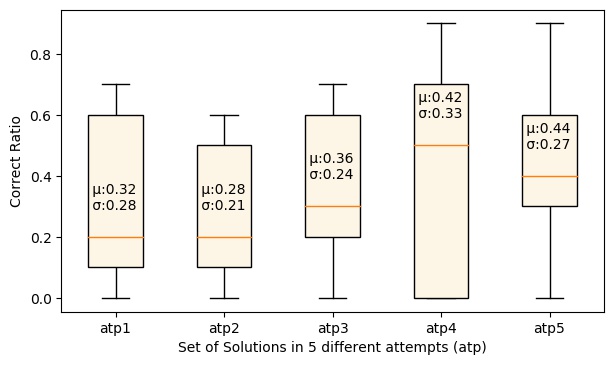}\label{fig:atp_cor}}
 \caption{\textbf{Evaluation of correct solutions generated by Copilot.} Plot (a) shows the normalized values for pass@Topk metrics against different values of k. It shows the fraction of correct solutions between Topk solutions of $5$ different attempts. Plot (b) shows the distribution, average and standard deviation of the Correct Ratio (CR) in each attempt for different programming tasks. }
  \label{fig:correct_ratio}
  \vspace{-1em}
\end{figure*}

Figure~\ref{fig:top_cor} shows the normalized values for pass@Topk of each programming task for Copilot. TopK solutions range between Top1 to Top10 because each attempt on Copilot includes only the Top10 suggestions. Based on this result, Copilot cannot find correct solutions for ``q2: Unique Dates Months''. This task asks for \textit{``...solve the problem by implementing 3 different functions...''}. Copilot could not understand this point within the task description and tried to solve the problem in one function. Thus, all of Copilot's solutions for this task failed the test cases because the test units of this task are based on implementing $3$ different functions. 

There are no correct solutions in Copilot's Top3 suggestions for ``q4: Sorting Tuples'' in $5$ different attempts. It increases to $0.02$ in the set of Top4 solutions. For ``q1'', ``q3'', and ``q5'', the pass@Top1 is equal to 0.08, 0.13, and 0.13, respectively. For some questions, the pass@Topk, at different values of k, shows greater values than the other questions. For example, ``q5'' has the greatest values for pass@Top4 and above. Also, ``q4'' has the lowest pass@Topk, for different values of k, after ``q2''.  

In general, pass@Topk increases by increasing the k. It means collecting a larger number of solutions suggested by Copilot increases the number of correct solutions and this growth can be different for different programming tasks.

In addition, Figure~\ref{fig:atp_cor} shows the Correct Ratio (CR) of solutions in each attempt independently. However, the distribution of CRs in different attempts is varied, but adding new attempts can increase the average CR of solutions. For example, the average CR in the first attempt (atp1) is equal to $0.32$ while it increases to $0.44$ in the last attempts (atp5). It shows if we ask Copilot to solve the same problem multiple times (here $5$ attempts), there is a chance to increase the CR among new Top10 suggested solutions on average. However, this is not correct for all questions. For example for ``q1'', the CR in ``atp4'' is $0.7$ but it decreases to $0.4$ in ``atp5''.  But, for ``q5'', the CR in the first attempt is equal to $0.7$ and it increases to $0.9$ in the last attempt.

Since we cannot calculate pass@Topk for students, in Table~\ref{tab:correct_ratio}, we compare the CR of solutions generated by Copilot with the CR of students' submissions. For this comparison, we calculate three different CRs for Copilot. The first, CR@Top1, reports the number of correct solutions out of all Top1 solutions in $5$ different attempts for each programming task. CR@Top5 calculates the fraction of correct solutions out of all Top5 solutions suggested by Copilot in $5$ different attempts. Finally, CR@Top10 represents the number of correct solutions generated by Copilot out of all its $50$ solutions for a programming task. Collecting more solutions decreases the CR of Copilot since it increases the fraction of wrong solutions. For some of the questions, CR@Top1 and CR@Top5 of Copilot are greater than students' CR. For all questions, the CR of students' submissions is greater than CR@Top10 for Copilot's suggestions. On average for all the programming tasks, the Correct Ratio (CR) of students' submissions is greater than the CR of Copilot's suggestions.

\begin{table}[t]
\centering
\noindent
\vspace{-5pt}
\caption{The Correct Ratio (CR) of Copilot's solutions while collecting Top1, Top5, and Top10 solutions in all $5$ attempts compared to the Correct Ratio (CR) of students' submissions}
\resizebox{\linewidth}{!}{
\begin{tabular}{c c c c c c}
\toprule
\multicolumn{2}{c }{\textbf{}} &\multicolumn{3}{c}{\textbf{Copilot}}& \multicolumn{1}{c}{\textbf{Students}} \\
\cmidrule(lr){1-2}
%\cmidrule(lr){2-2}
\cmidrule(lr){3-5}
\cmidrule(ll){6-6}

 & \textbf{Task}
      & \textbf{CR@Top1} & \textbf{CR@Top5} & 
     \textbf{CR@Top10} &  
     \textbf{CR}
     \\
      
\cmidrule(lr){1-2}
%\cmidrule(lr){2-2}
\cmidrule(lr){3-5}
\cmidrule(ll){6-6}

\textbf{q1} & \multicolumn{1}{ l }{\textbf{Sequential Search}} & \textbf{0.6} & 0.44 & 0.36 &	0.57 \\

\textbf{q2} &\multicolumn{1}{ l }{\textbf{Unique Dates Months}} &   0.00 & 0.00 & 0.00  &	\textbf{0.40}  \\

 \textbf{q3} &\multicolumn{1}{ l }{\textbf{Duplicate Elimination}} & \textbf{1} & 0.72 & 0.56 &	0.64 \\
 
\textbf{q4} &\multicolumn{1}{ l }{\textbf{Sorting Tuples}} &  0.00 & 0.08 & 0.14 &	\textbf{0.54} \\

\textbf{q5} & \multicolumn{1}{ l }{\textbf{Top-k Elements}} & \textbf{1} & 0.92 &  0.76  &	0.79 \\
\hline
  & \textbf{Total} &    0.52 &	0.43&	0.35&		\textbf{0.59} \\

\bottomrule
\end{tabular}
}
\label{tab:correct_ratio}
\end{table}

\subsubsection{\textbf{Repairing costs of Buggy solutions generated by Copilot and students}}\label{sec:repair_result}

In this part, we compare the repair cost of buggy solutions for Copilot with students. %\subsection{The Intersection between Correct and Buggy Solutions Generated by Copilot}
%As we already discussed in section-x\Amin{refer to a label please!},
As we already discussed, our observation shows there are buggy solutions that are generated by Copilot and are very similar to correct solutions. A small change can convert them into a correct solution.  Therefore, we attempt to quantify our observation by calculating the intersection between Copilot's correct and buggy solutions for each problem using the BLEU score~\cite{papineni2002bleu}. The comparison has been done in a pairwise manner between each correct and each buggy solution. For example, if out of $50$ solutions, $40$ are correct and $10$ are buggy, we end up with $400$ pairwise comparisons.

%\Amin{do we need all these details about BLEU?we can quickly explain it, refer to the some papers, and then report our results}
%BLEU is used in evaluating program synthesis approaches such as text-to-code, code summarization, and code prediction. %However, there are studies that show BLEU score is biased when applying on code~\cite{ren2020codebleu,tran2019does}. ~
BLEU is used in evaluating program synthesis approaches such as text-to-code, code summarization, and code prediction.  BLEU score uses the n-gram overlap between tokens of two contents and penalizes length difference. It returns a value between 0 and 1~\cite{tran2019does}. BLEU measures how well two texts match or are similar to each other. \citet{ren2020codebleu} introduces a new metric, called CodeBLEU, that measures the BLEU score on syntax and semantics of codes. As a part of this new metric, they measure CodeBLEU between AST of codes.

To measure the overlap between correct and buggy solutions, we measure the BLEU score between the AST of the buggy and correct. We omit those buggy codes which have syntax errors and cannot be converted into AST.  For example, the BLEU score of more than 0.7 between the AST of several correct and buggy pairs of solutions implies a high similarity between these two solutions. It can give us an estimation of the number of changes that we need to apply to a buggy solution to repair it.

\begin{figure*}
\centering
  \includegraphics[width=0.8\linewidth]{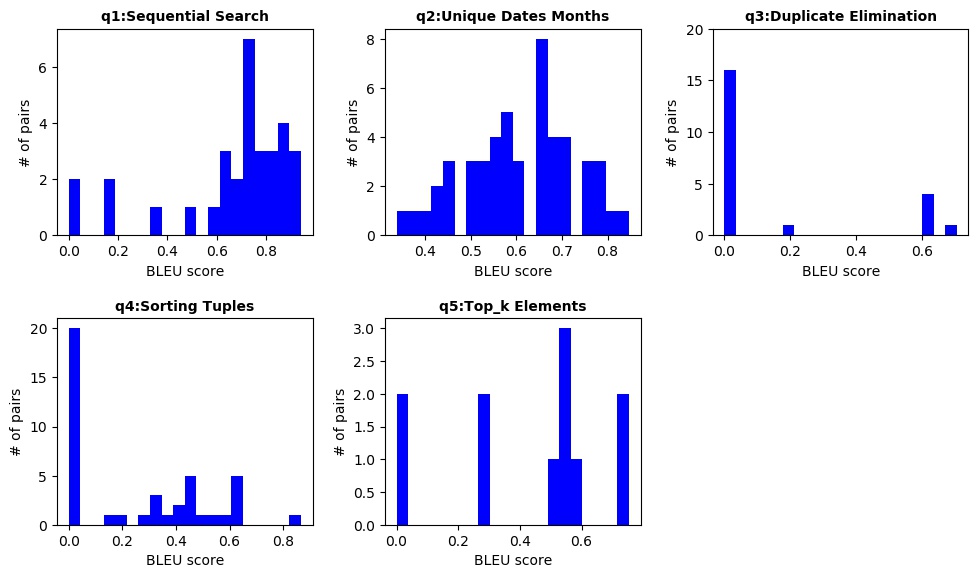}
  \caption{ \textbf{Distribution of BLEU score among the pair of correct and buggy solutions generated by Copilot.} This chart shows a histogram of the BLEU Score on pairs of correct and buggy solutions generated by Copilot. The BLEU score of $0.75$ and above represents a great similarity between the AST of a correct and buggy pair. The BLEU score between several pairs of the buggy and correct solutions is greater than $0.7$, in different programming tasks. This supports our observation that several buggy solutions can be corrected with small changes.}
  \label{fig:bleuscore_buggy}
  %\vspace{-5pt}
\end{figure*}

Figure~\ref{fig:bleuscore_buggy} shows the density distribution for the BLEU score among pairs of the buggy and correct solutions generated by Copilot for different programming tasks. As we can see in this figure, there are pairs of correct and buggy solutions with BLEU scores of 0.75 or greater. It shows that sometimes a small change in a buggy solution generated by Copilot can easily convert it into a correct solution, for example, changing ``$>$'' (greater) to ``$\geq$'' (greater equal). 

\begin{figure*}
\centering
  \includegraphics[width=0.8\textwidth]{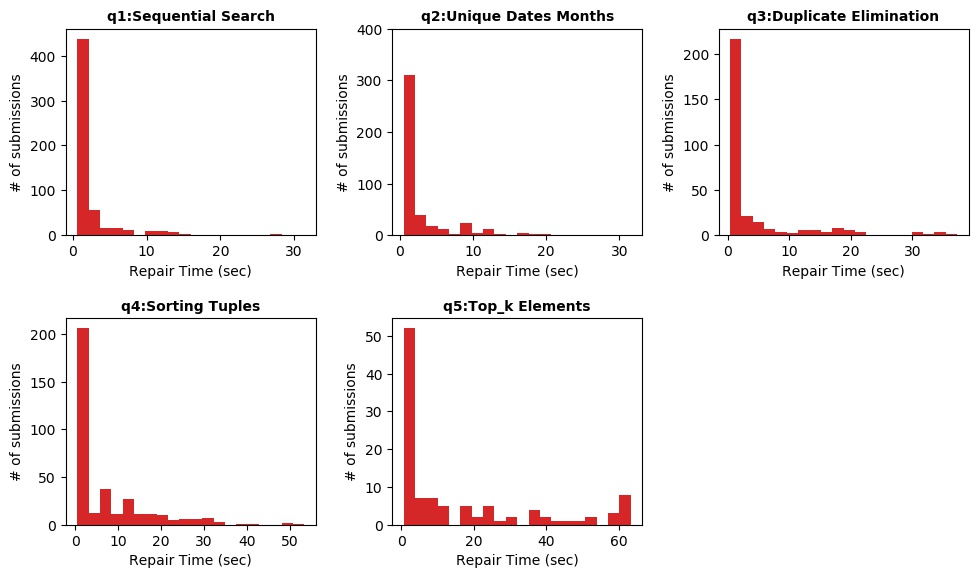}
  \caption{ \textbf{The distribution of repairing time for students' buggy submissions.} This chart shows a histogram of students’ buggy submissions based on their repairing time. It shows that there are more buggy submissions with low repairing time than buggy submissions with high repairing time. We repeat the downsampling process on students' submissions $5$ times to observe the same distribution in samplesets.}
  \label{fig:repairtime}
    \vspace{-5pt}
\end{figure*}

Now that some of the buggy solutions generated by Copilot are very similar to the correct solutions, we are interested in comparing the repairing cost of Copilot's buggy solutions with students' buggy submissions. As we have explained in Subsection~\ref{sec:resample}, for this comparison, we need to downsample students' submissions to the same size as Copilot's suggestions. Figure~\ref{fig:repairtime} shows the distribution of repairing time for repairing students' buggy submissions. There are a high number of submissions with low repairing time and few with high repairing time. Thus, to keep the distribution of repairing costs in the sample set close to the entire population, we repeat the downsampling process $5$ times and report all repairing metrics for students' submissions based on the average of all $5$ sampleset.

As we can find in Table\ref{tab:repair_result}, the average repair rate for Copilot's buggy solutions is greater than students', which are $0.95$ and $0.89$ respectively. This means that on average, $95\%$ of buggy solutions generated by Copilot have been fixed after the repair process. For example, for  ``q4: Sorting Tuples''  and ``q5: Top-k Elements'', all buggy solutions of Copilot ($100\%$) have been fixed while the repairing rate of students' submissions for these two tasks is equal to $85\%$.

In addition, the average repair time for Copilot's buggy solutions is less than the students'. This means that not only the repairing tool can fix the majority of Copilot's buggy solutions but also it can fix them faster than student buggy submissions. The average repairing time for Copilot's buggy solutions is $4.94$ seconds while it is equal to $6.48$ seconds for the students. The reason is that on average, the Relative Patch Size (RPS) of Copilot's buggy solutions that need to be repaired is smaller than students’. As we can find in Table~\ref{tab:repair_result}, the average RPS for Copilot and students are $0.33$ and $0.35$, respectively. 

%\Amin{this is fine, but can be used in conclusion or discussion as well}
We can conclude that however on average, the CR of students’ submissions is greater than Copilot's solutions, but the repairing costs of buggy solutions of Copilot are less than students. With a repairing tool, we can repair the majority of buggy solutions generated by Copilot and increase its CR.

 Thus, if Copilot, as a pair programmer in a software project, suggests buggy solutions, it is less expensive to fix its bugs compared to bugs that may be produced by junior developers when solving the same programming task.

\begin{table*}[htbp]
\centering
\noindent
%\vspace{-10pt}
\caption{Comparing the Repairing Cost of Copilot's suggestions with students's submissions}
\resizebox{0.7\textwidth}{!}{
\begin{tabular}{c c c c c c c c }
\toprule
\multicolumn{1}{c }{\textbf{}} & \multicolumn{1}{c }{\textbf{}} &\multicolumn{3}{c}{\textbf{Copilot}}& \multicolumn{3}{c}{\textbf{Students}} \\
\cmidrule(lr){1-2}
%\cmidrule(lr){2-1}
\cmidrule(lr){3-5}
\cmidrule(ll){6-8}

      &\multirow{-2}{*}{\textbf{Task}} &
     \shortstack[c]{\textbf{Rep}\\\textbf{Rate}} & 
      \shortstack[c]{\textbf{Avg Rep}\\\textbf{Time(sec)}} &  
      \shortstack[c]{\textbf{Avg}\\\textbf{rps}}  &   
     \shortstack[c]{\textbf{Rep}\\\textbf{Rate}} & 
      \shortstack[c]{\textbf{Avg Rep}\\\textbf{Time}} &  
      \shortstack[c]{\textbf{Avg}\\\textbf{RPS}}\\
      
\cmidrule(lr){1-2}
\cmidrule(lr){3-5}
\cmidrule(ll){6-8}

\textbf{q1} & \multicolumn{1}{ l }{\textbf{sequential search}} &	0.94 &	9.61 &	0.48 &	0.98 &	2.58 &	0.40\\

\textbf{q2}& \multicolumn{1}{ l }{\textbf{unique dates months}} &   	0.92 &	3.26 &	0.28 &	0.82 &	3.81 &	0.44 \\

 \textbf{q3} & \multicolumn{1}{ l }{\textbf{duplicate elimination}} &	0.91 &	0.64 &	0.26 &	0.96 &	4.35&	0.30\\
 
\textbf{q4} & \multicolumn{1}{ l }{\textbf{sorting tuples}} &   	\textbf{1.00} &	\textbf{0.78} &	\textbf{0.15} &	0.85 &	8.82&	0.29 \\

\textbf{q5}& \multicolumn{1}{ l }{ \textbf{top-k elements}} &       	\textbf{1.00} &	10.40&	0.50&	0.85&	12.84&	0.30 \\
\hline
& \textbf{Total} &    	\textbf{0.95}&	\textbf{4.94}&	\textbf{0.33}&	0.89&	6.48&	0.35\\

\bottomrule
\end{tabular}
}
\label{tab:repair_result}
\end{table*}

\subsubsection{\textbf{Diversity of Copilot's suggestions and students' submissions}}\label{sec:res:divers}

The diversity of solutions shows the novelty of Copilot and students in solving different problems. Also, it shows that while increasing the number of sample codes increases the fraction of correct solutions, this increment is due to the diversity of correct solutions or duplication. As we discussed in Subsection~\ref{sec:criteria}, we observe duplicate solutions in a single attempt and across multiple attempts on Copilot to solve a problem. On the other hand, we observe duplicate solutions among students' submissions as well. For example, for ``q1: Sequential Search'', after comparing the ASTs of students' correct submissions, $54.32\%$ of their submissions are identified to be duplicated.

To compare the diversity among students' submissions and Copilot's solutions, we randomly downsample $10$ student submissions in $5$ different sample sets and consider them as $5$ different attempts. Then, in each attempt on Copilot and for each sample set of students' submissions, we eliminate duplicate correct and buggy solutions. There are a few buggy solutions for Copilot and for student solutions involving syntax errors that cannot be converted into AST (3 solutions). We consider them as non-duplicate buggy solutions.

\begin{figure*}
\centering
  \includegraphics[width=1\textwidth]{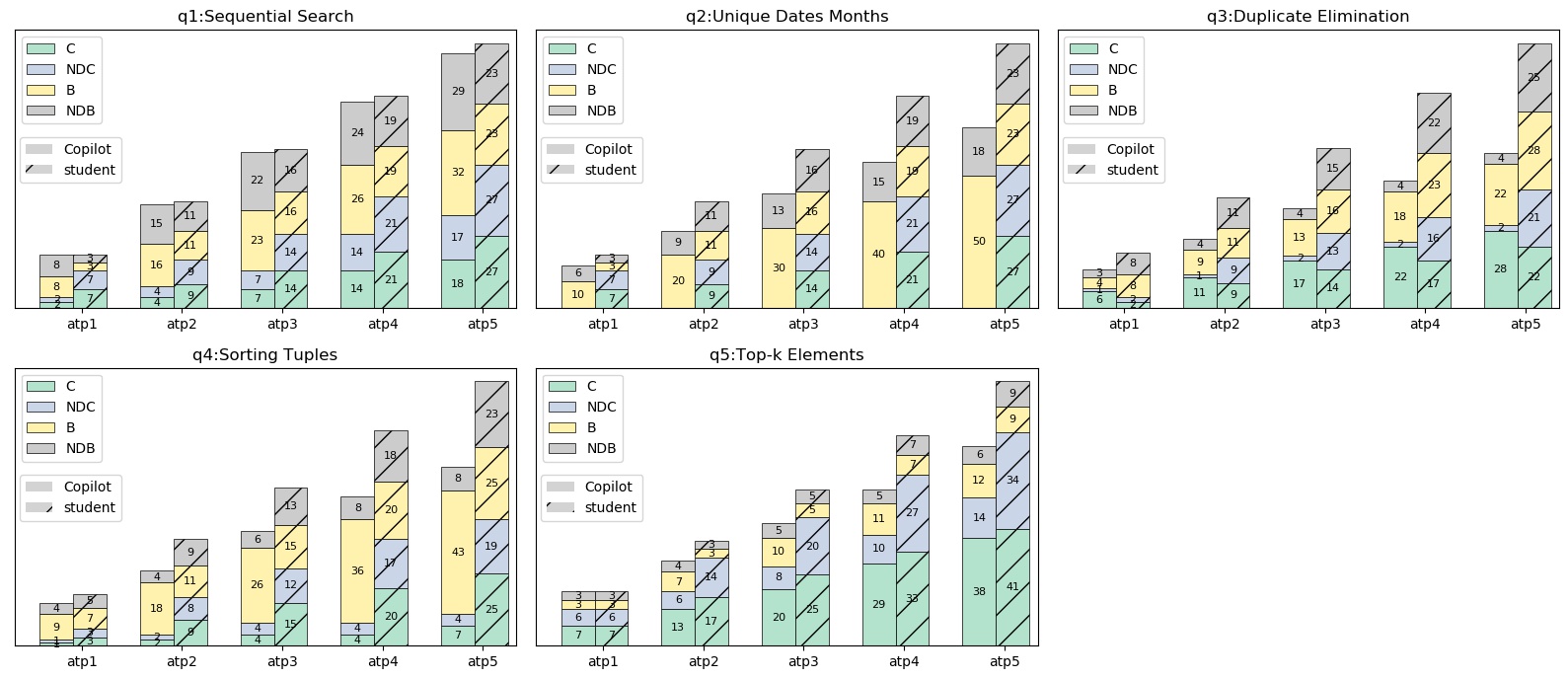}
  \caption{ \textbf{The cumulative distribution of solutions by Copilot and students.} It shows the cumulative distribution of Correct (C), Non-duplicate Correct (NDC), Buggy (B), and Non-duplicate Buggy (NDB) solutions for Copilot and students. Attempts (atp) for students equal to a random sampleset of their submission. Each value on the stack represents the number of solutions in each of the 4 categories. The growth of NDC solutions for Copilot's solutions decreases or stops for some programming tasks while the number of its Correct (C) solutions increases. Students’ submissions are more diverse than Copilot's solutions.}
  \label{fig:cumulativedist}
\end{figure*}

Figure~\ref{fig:cumulativedist} shows the cumulative distribution of Correct (C) solutions, None Duplicate Correct (NDC) solutions, Buggy (B) solutions, and None Duplicate Buggy (NDB) solutions by Copilot and students across different tasks.  In this figure, for example, in ``q3: atp3'', the number of Correct (C) solutions suggested by Copilot is 17 but the number of Non-duplicate Correct (NDC) solutions is only 2. This means that after generating more solutions and running more attempts, Copilot repeats these 2 correct solutions several times. However, out of 14 Correct (C) solutions generated by students in the third attempt (atp3), 13 solutions are non-duplicate. That is the same observation for buggy solutions. 
Increasing the number of attempts on Copilot leads to a jump in the number of correct solutions for ``'q1'' and ``q5'' from $2$ to $18$ and $7$ to $38$ respectively. However, for ``q3'' and ``q4'', this growth is smaller. The number of None Duplicate Correct (NDC) solutions of Copilot is less than or equal to the number of Correct (C) solutions in each attempt for each task. This is the same story for Buggy solutions. However, it shows that despite Copilot's claims that it removes the duplicate solutions, there are still duplicates in the Top 10 solutions of each attempt. 

The difference between C and NDC in student submissions is less than Copilot. For example, in ``q3'', the cumulative number of C solutions generated by Copilot in different attempts is greater than students' submissions in different samplesets. However, it is the opposite for NDC solutions. In ``atp5'' the cumulative number of C solutions generated by Copilot equals $28$ and it equals $22$ after the $5$ sampleset on students' submissions. However, the cumulative NDC solutions at these attempts equal $2$ (out of $28$) for Copilot and it equals $21$ (out of $22$) for students. It shows more diversity between correct and even buggy submissions of students compare to Copilot's solutions. 

As another example for Copilot, there is no more NDC solution after ``atp3'' for ``q3'' and ``q5''. This means that by increasing the number of solutions generated by Copilot for these two questions, the CR increases due to the duplication of correct solutions not generating new ones.

In general, the diversity of correct and buggy submissions for students is more than Copilot. While there is no guarantee that all non-duplicate solutions are optimized, students solved these 5 tasks with more diverse and novel solutions. 

\subsubsection{\textbf{The Cyclomatic Complexity of Codes}}
In this section, we calculate the Cyclomatic Complexity (C.C.) of codes generated by Copilot and students. Table~\ref{tab:Complexity} shows the average and the standard deviation of C.C. for the correct solutions generated by Copilot and students. It is worth mentioning that we use the sampling method explained in Subsection~\ref{sec:resample} to collect students' correct solutions. 

On average, the correct solutions suggested by Copilot are found to be more optimized than students' solutions. However, we should consider that for example, for ``q2'', Copilot has no correct solutions, or the CR of Copilot for ``q4'' is only $8\%$. In general, Copilot recommends less complex solutions than students for the same questions except for ``q1''. But, for ``q1'', the C.C. of Copilot's correct solutions have a lower standard deviation. It means that its C.C. is less spread around the average. Also, for ``q5'', Copilot used Python built-in functions ``Sort'' and ``Sorted'', however, it was asked in the description to not use them. 
\begin{table}[ht]
  \centering
  \caption{ The Cyclomatic Complexity (C.C.) of Copilot's solutions compare to students' submissions}
  \resizebox{\linewidth}{!}{
 \begin{tabular}{clcc}
%\begin{tabular}{c|l|c|c} 
\specialrule{.1em}{.05em}{.05em}

&\textbf{Question} & \textbf{C.C. Copilot} & \textbf{C.C. Students}
\\
%\midrule
\specialrule{.1em}{.05em}{.05em}
\textbf{q1}&{Sequential Search}& $5.8\pm 1.94$ &$4.63\pm 2.1$ \\
%\hline
\textbf{q2}&{unique dates Months} & - & $4.18\pm 1.03$\\
%\hline
\textbf{q3}&{Duplicate Elimination} & $3\pm 0.01$ & $3.12\pm 0.5$ \\
%\hline
\textbf{q4}&{Sorting Tuples} & $1\pm 0$ & $4.13\pm 1.03$\\
%\hline
\textbf{q5}&{Top\_k Elements} & $1.44\pm 0.69$ & $3.3\pm 1.46$\\

\specialrule{.1em}{.05em}{.05em}
&\textbf{Total} & $2.81$ & $3.87$\\

\specialrule{.1em}{.05em}{.05em}
\end{tabular}
}
  \label{tab:Complexity}
\end{table}

As already observed, we can conclude, the suggestions of Copilot can compete with students' solutions in C.C.

\subsubsection{\textbf{Syntactic Mastery}}
 As discussed in Subsection~\ref{sec:criteria}/(5), different developers can solve a programming task with different solutions. Consequently, this can impact the readability and maintainability of the code if it is not an efficient solution.

\begin{figure*}
\centering
  \includegraphics[width=0.8\linewidth]{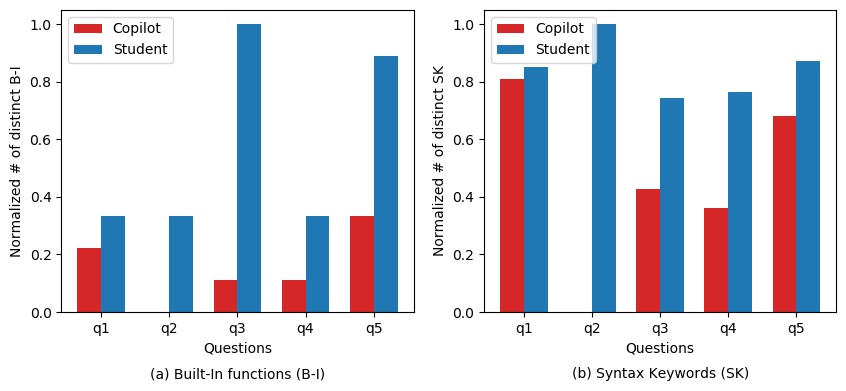}
  \caption{\textbf{R2-2: Diversity of programming Syntax Patterns in Solutions generated by Copilot and Students.} Plot (a) shows the normalized value for the distinct number of Python built-in functions in Copilot’s solutions compared to students’ for different questions. Plot (b) shows the normalized value for the distinct number of Python Syntax keywords in Copilot’s solutions compared to students.}
  \label{fig:pythonic}
  \vspace{-5pt}
\end{figure*}

In this section, we compare the diversity of syntax keywords and the usage of built-in functions between the solutions generated by Copilot and those written by humans for different programming tasks. Figure~\ref{fig:pythonic} shows the diversity of syntax keywords and built-ins that we observed in both Copilot's and students' solutions with normalized values. Students used more diverse keywords and built-ins in comparison to Copilot.

For example, for q3: Duplicate Elimination, the only Python built-in function in Copilot’s solutions is "append". However, students included more diverse built-ins such as \{'count', 'remove', 'index', 'copy', 'append', 'reverse', 'pop'\} in their solutions. As another example, in q5: Top-k elements, Copilot used \{'sort', 'append', 'remove'\} as built-in functions in all of its solutions but students used \{'copy', 'pop', 'remove', 'append', 'sort', 'extend', 'reverse', 'clear'\}. The using of programming keywords by Copilot and students is similar to built-ins. For example, for q4: Sorting Tuples, there are solutions provided by students that iterate over the list of tuples to sort them causing diverse syntax patterns in their solutions such as \{'Tuple', 'Lt',  'Add', 'Expr', 'Continue', 'Eq', 'Break', 'Gt', 'BoolOp', 'And', 'UnaryOp', 'USub', 'LtE'\}. We cannot find these programming patterns in Copilot's solutions as it only used the built-in function "sort" in the majority of its solutions.

Students used more diverse syntax patterns and built-ins to solve the same problem compared to Copilot. This may be the result of students not being familiar with advanced Python features as opposed to Copilot which uses such features frequently. However, this diversity could stem from the diversity of student submissions as discussed in Subsection~\ref{sec:res:divers}, or it could be the result of restriction in some assignments' descriptions, for example in q5: Top\_k elements that not using the built-in functions sort and sorted is requested which, unlike the students, Copilot was not able to understand this restriction. 

\begin{tcolorbox}[colback=blue!5,colframe=blue!40!black]
\textbf{Findings:} In general, Copilot suggests solutions that compete with students' submissions in different aspects. The correct ratio and diversity of students' submissions are greater than Copilot’s. However, the cost of repairing buggy solutions generated by Copilot is less than students'. In addition, the complexity of Copilot’s generated codes is less than students'.  \\
\textbf{Challenges:} Copilot has difficulty understanding some requirements in the description of tasks. This affects the correct ratio of its solutions. However, students understand those details and consider them in their submissions. 

\end{tcolorbox}

%%%%%%%%%%%%%%%%%%%%%%%%%%%%%%%%%%%%%%
\section{Discussion and Limitation}\label{sec:discussion}
%\Amin{I revised this section, have a look please!}
In this section, we discuss the boundaries of Copilot and how to make it more beneficial in real programming tasks despite its limitations.

\subsection{Description of Problems (Prompts)}

Our results show that Copilot cannot understand some details in the description of problems that are understandable by humans. For example, in q5, ``Top-k Elements'', it is asked in the description to ``... not use Python's built-in functions sort and sorted ...''. Copilot cannot understand this detail and uses these two built-in functions in all of the correct solutions. However, the majority of students avoided using these built-in functions. Instead, they wrote a sorting algorithm and then called it for sorting tasks or used other built-in functions such as ``append'', ``remove'' and ``max'' in their solutions. As our results in Subsection \ref{sec:algorithms} show, Copilot suggests correct solutions for different sorting algorithms (meaning that Copilot is familiar with different sorting algorithms such as ``Bubble Sort'' or ``Merge Sort''), but it did not use them in q5 because it could not figure out the requirements of the problem. But students apply their knowledge about sorting algorithms to solve this problem. Thus, since in the prompt, we cannot limit Copilot to NOT using certain functions, instead, it is better to clarify our task by defining functions that it is allowed to use.

In q4, ``Sorting Tuple'', it is asked to return the list of tuples in an order that ``... older people are at the front ...”. Copilot cannot understand this part. In $92\%$ of suggestions, it returned the sorted tuples in the default order: \textit{ascending}. However, students considered this point in their submission. We even checked some of the buggy submissions by students. Our observations show that even in buggy submission, students considered the correct order of sorting. It means that they fully understood what the point of sorting tuples is in a way that ``...older people are at the front...''.

Copilot shows similar limitations on algorithmic problems. For example, when asking Copilot to implement the ``activity'' class in Subsection \ref{subsec:ga}, Copilot cannot understand putting limits on variables even though it was asked to do so explicitly. Another limitation is its difficulties in understanding long descriptions which are also observed by \cite{li2022competition}. Throughout our testing in Subsections \ref{sec:algorithms} and \ref{sec:humans}, we observed that Copilot might misunderstand the problem entirely if the description contains multiple sentences (whether short or long).

\subsection{Experimental Suggestions}
Furthermore, for more exploration on how to change prompt to meet the target solution, we performed some experiments by applying different scenarios and discussing their impacts on the results.

\textbf{\textit{Scenario\#1:}} In this scenario, we changed ``...older people are at the front...'' to ``...descending order...'' in the description of q4 and repeated the process with Copilot to generate solutions. This small change improves the CR from $14\%$ to $79\%$. This improvement shows there are some details/keywords in the description of problems that seem obvious to humans, but Copilot cannot understand those details in natural language. If we change those details into programming specific/technical keywords such as ``descending'', it can help Copilot recommend relevant solutions.\\
 
\textbf{\textit{Scenario\#2:}} We have a similar observation for q2, ``Unique Birthday'', where the Copilot cannot understand the requirements mentioned in the description, however, all students considered it. In this question, it is asked for ``...implement 3 different functions unique\_day, unique\_month and contains\_unique\_day...'', to address the problem. Copilot could not understand this condition. Unit tests for q2 are testing all 3 functions. Thus, the CR of Copilot for q2 equals zero because all $50$ solutions in different attempts have failed on some of the test units. 

So, in this scenario, we gave 3 separate descriptions to Copilot for unique\_day, unique\_month, and contains\_unique \_day functions in the same source file. Here is the revised description that we used:
 
\begin{itemize}
    \item \textbf{unique\_day:} Given a day and a list of possible birthday dates %\Amin{dates?}, 
    return True if there is only one possible birthday with that day, and False otherwise.
    
    \item \textbf{unique\_month:} Given a month and a list of possible birthday dates, return True if there is only one possible birthday within that month, and False otherwise.
    
    \item \textbf{contains\_unique\_day:} Given a month and a list of possible birthday dates, return True if there is only one possible birthday with that month and day, and False otherwise.
\end{itemize}
 
We start with the description of unique\_day at the first line of the source file. Then, we accepted the first solution suggested by Copilot. We continued with the description of unique\_month in the next line and accepted the first suggested solution and followed the same instruction for contains\_unique\_day. We repeat the process 50 times to generate 50 solutions that contain 3 separate functions. Copilot even calls the unique\_day function in some of its suggestions for the contains\_unique\_day function. You can find sample solutions in the replication package. %Appendix-x\Amin{? or  replication package?}.
Since there are separate unit tests to test each function separately, we run related tests against each function. In this scenario, the CR of unique\_day, unique\_month, and contains\_unique\_day are $88\%$, $0\%$, and $40\%$ respectively.  

While the original description was clear to students, Copilot could not understand it. Instead of asking Copilot to solve the problem with different functions, we divide a problem into 3 different problems. It increases the CR for unique\_day and contains\_unique\_day. However, the CR of unique\_month is still zero. In the following, we investigate this case with a different scenario.\\

\textbf{\textit{Scenario\#3:}} Since Copilot could not find any correct solutions for unique\_month, we manually checked its suggested solutions. We found that in all buggy solutions, Copilot refers to the second item of the ``birthday'' tuple in the list of birthday dates as the month. However, unit tests consider month as the first item of tuples to test the functionality of the method. For example, consider below unit test:
\begin{itemize}
    \item unique\_month (Month =  ``January'', Birthdays = \newline [( ``January'',``1'' ), ( ``January'', ``2'' )]).
\end{itemize}
 
In each tuple in the list of birthdays, for example, (``January'',``1''), Copilot referred to the second item as a month, however, the first item in the tuple is the birthday month. 
 
In the description of ``unique\_month'', we added the above unit test as a sample input, at the end of the description. It improves the CR of ``unique\_month'' from $0\%$ to $91\%$. It shows that adding sample input or sample unit test in the description of problems can help Copilot to generate more correct solutions. 

In addition, we randomly checked $20\%$ of students' submissions (both correct and buggy). Our observation shows that none of them assumed any wrong structure for the input data, while the structure of input is not clear in the description of the question. Thus, we assume that there is some extra clarification between students and the lecturer about the structure of the input.

%Our observation shows Copilot tries to mimic the developer's naming, commenting, and coding style. As explained and laid out in our replication package \cite{GithubRepo}, we used comments and doc-strings to separate Copilot's output from our own notes. We also wrote some unit testing methods for each script to make sure of the functionality of each script. After some time, we saw that Copilot was generating similar testing snippets and comments at the end of some suggestions for the same problem.

% It is also worth mentioning that if Copilot can understand the problem at hand, it can output code that is follows a different execution logic than what was requested. However, the generated code is correct and solves the problem. As an example, during testing Copilot for generating solutions to the selection sort algorithm, our definition of the problem followed \cite{cormen2022introduction}'s. Our description of the problem, required Copilot to generate code in a specific manner. However, Copilot's generated code solved the problem correctly but with a different solution which was close to the codes for this problem found on the web.    
\section{Threats to Validity}\label{sec:threats}

 We now discuss the threats to the validity of our study following the guidelines provided by Wohlin ~\cite{wohlin2012experimentation} for experimentation in software engineering.

\subsection{Internal Validity}
The threat to internal validity comes from the fact that Copilot is closed-source. We cannot analyze our results based on the characteristics (and expected behavior) of Copilot’s trained model. This is also the case for Copilot’s training data, hence we are not able to indicate whether it memorized the solutions to these inquiries from its training set or whether it generates a unique solution. Similar to other researchers~\cite{NguyenMSR22, imai2022github, vaithilingam2022expectation, chen2021evaluating}, we can only investigate Copilot’s functionality in suggesting code for the provided prompt.

Also, as our experiments have shown, Copilot's suggestions change over time and are not always consistent. This may come from the inconsistency stemming from the nature of LLMs and also the continuous improvement of Copilot's engine as an ML product, perhaps by feeding new code samples or learning from new queries submitted to Copilot. As a result, we cannot guarantee that other researchers will receive the same suggestions and results that we obtained by performing the same experiments.

\subsection{External Validity}
The lack of a dataset that comes from an industrial context and contains programming task statements along with their corresponding codes drives us to follow the path of other research in software engineering using classical programming tasks to study Copilot's competence~\cite{vaithilingam2022expectation, sobania2021choose, NguyenMSR22, drori2021solving, tang2021solving}.  There are different advantages to these types of programming tasks that we discussed in Subsections~\ref{sec:RQ1_all} and~\ref{sec:dataset}. To highlight two advantages, first, Copilot is able to generate answers corresponding to these task descriptions. Thus, we could apply our assessments beyond the correctness of the suggested solutions. Also, the task descriptions in our datasets are human-written and it decreases the possibility of the memorization issue in LLMs. But these programming tasks are not representative of the whole programming tasks in real software projects. 

Considering the choice of programming tasks and to have a fair comparison, we compared Copilot with students in a Python programming course. While we have no information about the background and characteristics of the participants, we assume that they are good representatives of junior developers in real software projects, but they may not be representatives of the whole population.

\subsection{Conclusion Validity}
To mitigate the threats to the validity of our conclusions, we choose different quantitative metrics, based on other studies in software engineering, to compare Copilot’s codes with humans’~\cite{fakhoury2019improving, NguyenMSR22, kim2006long}. Even though these quantitative metrics reduce the chance of having biased conclusions, they do not enable us to conduct any qualitative assessment such as how humans interact with the tool.

\subsection{Construct Validity}
The threat to the construct validity of our study stems from the fact that all the features and capacities of a good AI pair programmer cannot be captured by quantitative metrics. Since pair programming is an interaction between human-human or human-tool, the opinion of humans about their experience in using such tools as an AI pair programmer is also required for a comprehensive study. For example, someone may prefer a pair programming tool that accepts voice commands to a tool that suggests a list of possible solutions because they like the discussion part of pair programming more than seeing a list of suggestions.

\section{Conclusion}\label{sec:conclusion}
In this paper, we have studied Copilot's ability on code generation and compared its generated codes with those of humans. Our results show that Copilot is %For testing its ability to generate codes for fundamental algorithms which are used in software development, our results demonstrate that Copilot is
able to generate correct and optimal solutions for some fundamental problems in algorithm design. %understand the problems and output correct, optimal codes for them.
However, the quality of the generated codes depends greatly on the conciseness and depth of the prompt that is provided by the developer. Furthermore, our results indicate that Copilot still needs more development in fully understanding natural language utterances in order to be able to fill-in the position of a pair-programmer. Copilot may occasionally be unable to generate code that satisfies all the criteria described in the prompt, but the generated code can be incorporated with little to moderate changes to the provided prompt or the code. %Our results by comparing Copilot's generated codes with those of humans also reinforce this finding.

 Although, Copilot suggests solutions that are more advanced in programming than solutions provided by junior developers, and even though those solutions are comparable to humans’ in correctness, optimality, reproducibility, and repairing costs, an expert developer is still required to detect and filter its buggy or non-optimal solutions. Thus, Copilot can be an asset in software projects if it is used by expert developers as a pair programmer but it can turn into a liability if it is used by those who are not familiar with the problem context and correct coding methods. 
 
Given that Copilot has recently been released as a commercial product, a new wave of developers will have access to it. This will undoubtedly enrich Copilot's training dataset and will also expose more of its shortcomings. However, Copilot solutions can be troublesome if novice developers/students fully trust them, on the other hand, we hypothesize that Copilot’s suggestions may help them in improving their programming skills. Therefore, as future work, a tool or a layer on top of Copilot that can filter out buggy and non-optimal suggestions will reduce the liability of using this tool in software projects. Future works can also use our study design and explore more diverse programming tasks with heterogeneous participants in a human-centered study, to more comprehensively compare Copilot with humans as an AI pair programmer.
%Also, we suggest investigating the current RQs further, by exploring more diverse programming tasks, in a human-centered study, to indicate the capabilities of Copilot as an effective programming assistant for humans.

%Moreover, our results show that even though Copilot may be unable to generate codes that satisfy all the criteria described in the prompt, the generated codes can be incorporated by the developer with little to moderate change to the provided prompt or the generated codes.
% We suggest future works to focus on exploring more diverse programming tasks, in a human-centered study, to indicate the capabilities of Copilot as an effective programming assistant for humans.
%Given that recently Copilot has been released as a commercial product, a new wave of developers will have access to it. This will undoubtedly increase Copilot's training dataset and will also expose more of its shortcomings. However, both outcomes may result in an improvement of Copilot's ability as pair-programmer over time. Therefore, for future work, we suggest investigating the current RQs further, by exploring more diverse programming tasks, in a human-centered study, to indicate the capabilities of Copilot as an effective programming assistant for humans.
%\Foutse{we should add some recommendations for future works...}
%\section*{Acknowledgements}
%This work is partly funded by the Natural Sciences and Engineering Research Council of Canada (NSERC), the Fonds de Recherche du Québec (FRQ), and Institut de VAlorisation des DOnnées (IVADO).\Amin{@Foutse, could you please check this?}
%\balance
\bibliography{mybibfile}
%\end{sloppypar}
\end{document}